\documentclass{aa}
\usepackage{graphicx}
\usepackage{txfonts}
\usepackage{aalongtable}
 \usepackage{newclude}
\usepackage[round]{natbib}
\bibpunct{(}{)}{;}{a}{}{,}
\graphicspath{{./images/}}
\begin{document}

\title{The planetary nebula population in the halo of M87}

\author{A. Longobardi\inst{1}, M. Arnaboldi\inst{2},
  O. Gerhard\inst{1}, L. Coccato\inst{2}, S. Okamura\inst{3},
  K.C. Freeman\inst{4}}

\offprints{A. Longobardi}

\institute{Max-Planck-Institut f\"ur Extraterrestrische Physik,
  Giessenbachstrasse, D-85741 Garching, Germany \\
  e-mail: alongobardi@mpe.mpg.de, gerhard@mpe.mpg.de \and European
  Southern Observatory, Karl-Schwarzschild-Strasse 2,
  D-85748 Garching, Germany \\
  e-mail: marnabol@eso.org \and Department of Advanced Sciences,
  Faculty of Science and
  Engineering, Hosei University, Tokyo, 184-8584 Japan\\
  e-mail: sadanori.okamura@hosei.ac.jp
  \and RSAA, Mt. Stromlo Observatory, Australia\\
  e-mail:kcf@mso.anu.edu.au}

\date{Accepted 15.08.2013}

   \authorrunning{A.Longobardi et al.}

 
\abstract 
{} 
{We investigate the diffuse light in the outer regions of the nearby
  elliptical galaxy M87 in the Virgo cluster, in the transition region
  between galaxy halo and intracluster light (ICL).}
{The diffuse light is traced using planetary nebulas (PNs). The
  surveyed areas are imaged with a narrow-band filter centred on the
  redshifted [OIII]$\lambda 5007$ \AA\ emission line at the Virgo
  cluster distance (the on-band image) and with a broad-band V-filter
  (the off-band image).  All PNs are identified through the on-off
  band technique using automatic selection criteria based on the
  distribution of the detected sources in the colour-magnitude diagram
  and the properties of their point-spread function.}
 {We present the results of an imaging survey for PNs within a total
   effective area of 0.43 $\deg^2$, covering the stellar halo of M87
   up to a radial distance of 150 kpc. We extract a catalogue of 688
   objects down to $m_{5007}=28.4$, with an estimated residual
   contamination from foreground stars and background Ly$\alpha$
   galaxies, which amounts to $\sim 35\%$ of the sample. This is one of
   the largest extragalactic PN samples in number of candidates,
   magnitude depth, and radial extent, which allows us to carry out an
   unprecedented photometric study of the PN population in the outer
   regions of M87. We find that the logarithmic density profile of the
   PN distribution is shallower than the surface brightness profile at
   large radii. This behaviour is consistent with a model where the
   luminosity specific PN numbers for the M87 halo and ICL are
   different. Because of the depth of this survey we are also able to
   study the shape of the PN luminosity function (PNLF) in the outer
   regions of M87. We find a slope for the PNLF that is steeper at
   fainter magnitudes than the standard analytical PNLF formula and
   adopt a generalised model that treats the slope as a free
   parameter.}
   {The logarithmic PN number density profile is consistent with the
     superposition of two components associated with the halo of M87
     and with the ICL, which have different $\alpha$ parameters. We
     derive $ \alpha_{2.5,\mathrm{halo}}=(1.10^{+0.17}_{-0.21})\times
     10^{-8}$ N$_{\mathrm{PN}}$L$^{-1}_{\odot,\mathrm{bol}}$ and
     $\alpha_{2.5,\mathrm{ICL}}=(3.29^{+0.60}_{-0.72})\times10^{-8}$
     N$_{\mathrm{PN}}$L$^{-1}_{\odot,\mathrm{bol}}$ for the halo and
     the intracluster stellar components, respectively. The fit of the
     generalised formula to the empirical PNLF for the M87 halo
     returns a value for the slope of $1.17$ and a preliminary
     distance modulus to the M87 halo of $30.74$. Comparing the PNLF
     of M87 and the M31 bulge, both normalised by the sampled
     luminosity, the M87 PNLF contains fewer bright PNs and
     has a steeper slope towards fainter magnitudes.}


   \maketitle
%

 \section{Introduction}

Stars in the mass range between $1$ and 8 M$_{\odot}$ go through the
planetary nebula (PN) phase before ending their lives as white dwarfs.
The optical image of a PN is dominated by the luminous ionized
envelope that is powered by the stellar core at its centre. The
envelope emits in several strong lines from the UV to the NIR, and
\citet{dopita92} showed that up to 15\% of the luminosity of the
central star is re-emitted in the forbidden [OIII] line at
$\lambda$5007 \AA.

PNs have been used as kinematic tracers of the stellar orbital
distribution in the outer regions of galaxies, where the continuum
from the stellar surface brightness is too low with respect to the
night sky, both in nearby galaxies \citep{hui93,peng04,merrett06,
  coccato09,cortesi13} and out to 50-100 Mpc
\citep{ventimiglia11,gerhard05}.  The outer regions of galaxies are
particularly interesting because dynamical times are longer there, so
they may preserve information of the mass assembly processes.

The observed properties of the PN population in external galaxies
also correlate with the age and metallicity of the parent stellar
population.  These properties are the luminosity-specific PN number,
$\alpha$-parameter for short, that quantifies the stellar luminosity
associated with a detected PN, and the shape of the PN luminosity
function (PNLF). We discuss these in turn.

Observationally the values of $\alpha$ correlate with the integrated
(B-V) colour of the parent stellar population, with the spread in
observed values for the reddest galaxies increasing significantly with
respect to the constant value observed in bluer $((B-V) < 0.8)$
objects \citep{hui93,buzzoni06}. For the reddest galaxies, the value
of $\alpha$ correlates with the (FUV-V) colour, with the lowest number
of PNs observed in old and metal-rich systems \citep{buzzoni06}.
 
To describe the shape of the PNLF for extragalactic PN populations,
the analytical formula proposed by \citet{ciardullo89} has generally
been used, but see also models by \citet{mendez93,mendez08}. At the
brightest magnitudes the PNLF shows a cutoff that is observed to be
invariant between different Hubble types and has been used as
secondary distance indicator \citep{ciardullo04}. From about one
magnitude fainter than the PNLF cutoff, the analytical formula
predicts an exponential increase, in agreement with the slow PN fading
rate described by \citet{henize63}. Observationally, the PNLF slope
correlates with the star formation history of the parent stellar
population, with steeper slopes observed in older stellar populations
and flat or slightly decreasing slopes in younger populations
\citep{ciardullo04,ciardullo10}.

The Virgo cluster, its elliptical galaxies, and intracluster light
(ICL) were the targets of several PN surveys
\citep{ciardullo98,feldmeier98,feldmeier04,arnaboldi02,arnaboldi03,arnaboldi04,castro03,castro09,aguerri05},
aimed at measuring distances and the ICL spatial distribution. In this
paper we report the results of a deep survey carried out with the
Suprime-Cam at the Subaru telescope to study the PN population in the
halo around M87, one of the two brightest galaxies in the Virgo cluster.

NGC~4486 (M87) is a giant elliptical galaxy situated at the centre of
the subcluster A in the Virgo cluster \citep{binggeli87}, the nearest
large scale structure in the local universe. According to current
models of structure formation, M87 acquired its mass over a long
period of time through galaxy mergers and mass accretion. The stars in
M87 are old \citep{liu05}, and its stellar halo contains about 70\% of
the galaxy's light down to $\mu_{V}\sim27.0$ mag arcsec$^{-2}$
\citep{kormendy09}. Thus M87 is an ideal target for a survey aimed at
detecting PNs in an extended galaxy halo, when investigating the
halo kinematics and stellar population.

The paper is organized as follows: in Section~\ref{section2} we
describe the Suprime-Cam PN survey and the data reduction
procedure. In Section~\ref{section3} we describe the PN catalogue
extraction and validation.  The relation between the spatial
distribution of the PN candidates and the M87 surface brightness
profile is investigated in Section~\ref{section4}.  We present the
PNLF measured in the M87 halo in Section~\ref{section5} and discuss
the comparison with the PNLF for the M31 bulge. Finally, we summarize
our conclusions in Section~\ref{section6}. In the rest of the paper we
adopt a distance modulus of 30.8 for M87, which means that the
physical scale is 73 pc arcsec$^{-1}$.

 \section{The Suprime-Cam M87 PN Survey}\label{section2}
\subsection{Imaging and observations}\label{section2.1}

\begin{figure} \centering
   \includegraphics[width=9.4cm]{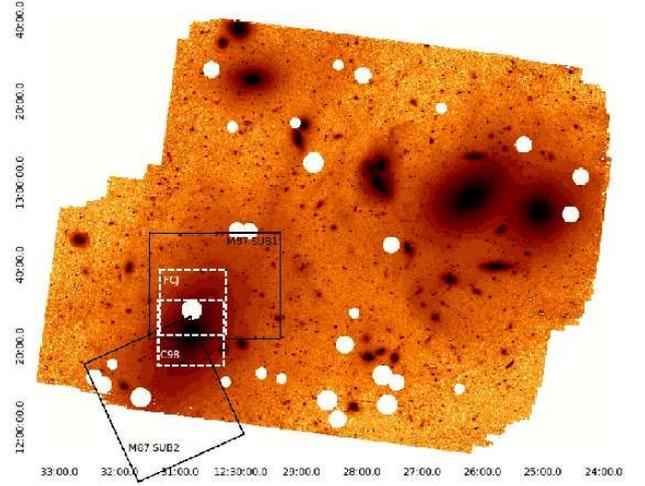}
   \caption{The core of the Virgo cluster \citep{mihos05} with the
     positions of the fields studied in this work (black
       rectangles) and in previous surveys by \citet{ciardullo98} and
       \citet{feldmeier03} (dotted white squares). The region is
     dominated by the halos of 3 bright galaxies, M84, M86 and M87,
     with M87 covered by the Suprime-Cam fields. North is up, East is
     to the left.}\label{mapMihos}
 \end{figure}

In March 2010 we observed two fields with the Suprime-Cam
10k$\times$8k mosaic camera, at the prime focus of the 8.2 m Subaru
telescope \citep{miyazaki02}. The CCDs have a readout noise of 10
e$^{-}$ and an average gain of 3.1 e$^{-}$ ADU$^{-1}$.  Each field of
view covers an area of 34$\arcmin$ $\times$ 27$\arcmin$, with a pixel
size of 0$\arcsec$.2; the two pointings cover the halo of M87 out to a
radial distance of 150 kpc.  Figure~\ref{mapMihos} shows a deep V-band
image of the Virgo cluster core region and the two fields studied in
this work overlaid.  We label these fields as the M87~SUB1 and
M87~SUB2 fields, respectively.

Both fields were observed through an [OIII] narrow-band filter
(on-band filter), centred on $ \lambda_{c}$=5029 \AA\ with a band
width $\Delta\lambda$=74 \AA and a broad-band V-filter (off-band
filter).  The total exposure time for the on-band images was $\sim3.7$
hr and $\sim4.3$ hr, while for the off-band images it was $1$ hr and
$1.4$ hr, for the M87~SUB1 and M87~SUB2 fields, respectively. Deep
V-band images are needed for the colour selection of the PN candidates
(see Sect.  \ref{section3}).

Our strategy for data acquisition was set up to achieve the best image
quality. Narrow-band and broad-band images were taken close to each
other during the observing nights to secure similar conditions in
terms of scattered light and atmospheric seeing, while calibration
images, such as dark sky\footnote{Dark sky is an image of an empty
  field (off target) on the sky}, were taken in between the science
images to have a similar S/N and image quality. 

The nights were photometric with an overall seeing on the images less
than 1$\arcsec$.  Airmasses were 1.01 and 1.06 for the reference
M87~SUB1 [OIII] and V-band exposures, and 1.13 and 1.03 for the
reference M87~SUB2 [OIII] and V-band exposures\footnote{The reference
  [OIII] and V-band exposures are the one used as reference images for
  the final image combination.}. We did not perform any measurements
for the extinction coefficients. We adopted the mean value of X=0.12
mag/airmass, as listed for the [OIII] and V filters at the Mauna Kea
summit web site
(http://www.jach.hawaii.edu/UKIRT/astronomy/exts.html), and in
agreement with \citet{buton13}.

The on-band filter bandpass was designed such that its central
wavelength coincides with the redshifted $\lambda$5007 \AA\ emission
at the Virgo cluster. The Johnson V-band filter can be used as the
off-band filter despite the fact that it contains the [OIII] line in
its large bandpass ($\sim1000$ \AA).  The depth of the Suprime-Cam
survey was chosen to detect all PNs brighter than $m^*$+2.5, where
$m^*$ is the [OIII]5007 \AA\ apparent magnitude of the bright cutoff
of the PNLF for a distance modulus 30.8.  Table \ref{table:1} gives a
summary of the field positions, filter characteristics and exposure
times for the on-band and off-band exposures for the analysed area
around M87.

%

%
 \begin{table*}
   \caption{Summary of the field positions, filter characteristics,
     exposure times, and seeing for the narrow-band (on-band) and
     broadband (off-band) images.}
   \label{table:1} 
   \centering
   \begin{tabular}[h!]{c c c c c c c c c c c c c}
     \hline\hline\\
     & & &  \multicolumn{4}{c}{[OIII] Filter} & & \multicolumn{4}{c}{V-band Filter} \\
     \cline{4-7} \cline{9-12} \\ 
     Field  & $\alpha$(J2000)  & $\delta$(J2000)&  $\lambda_{c}$ &  FWHM &
     Exposure  &   S$_{FWHM}$&  &  $\lambda_{c}$   &  FWHM  &   Exposure  &
     S$_{FWHM}$\\
     \hline 
     & (hh mm ss)& ($^{o}$ $\arcmin$  $\arcsec$) & (\AA\ ) & (\AA\ )
     & (s) & ($\arcsec$) & &( \AA\ ) & (\AA\ ) & (s) & ($\arcsec$) \\
     \hline \\
     
     M87 SUB1 &12 30 25.230 &+12 35 03.85 &5029 &74 &11 $\times$1200
     &0.96 &&5500 &956 &13$\times$360 &1.00 \\
     M87 SUB2 &12 31 15.581 &+12 07 22.33 &5029 &74 &10 $\times$1200 
     &0.98 &&5500 &956 &14 $\times$360 &0.8 \\
     \\
     \hline                  
   \end{tabular}
 \end{table*}

\subsection{Data Reduction, Astrometry and Flux Calibration}
\label{section2.2}

The removal of instrumental signature, geometric distortion
correction, background (sky plus galaxy) subtraction, a first
astrometric solution and the final image combination were done using
standard data reduction packages developed for Suprime-Cam data
(sdfred software v2.0, from
http://www.naoj.org/Observing/Instruments/SCam/sdfred).  Cosmic rays
identification and rejection were carried out with L.A.Cosmic
(Laplacian Cosmic Ray Identification) algorithm \citep{vandokkum01}.
Because of the spectroscopic follow-up our aim was to derive accurate
positions of the PN candidates.  To do this we improved the astrometry
of the images to get a relative positional accuracy with less than
$0\arcsec .3$ in the residuals.  The astrometric solution was computed
using image astrometry tasks in the IRAF\footnote{IRAF is distributed
  by the National Optical Astronomy Observatory, which is operated by
  the Association of Universities for Research in Astronomy (AURA)
  under cooperative agreement with the National Science Foundation.}
package $\bf{imcoords}$, performing a geometrical transformation by
2nd order polynomial fitting.  All images analysed in this paper are
on the astrometric reference frame of the 2MASS catalogue\footnote{The
  astrometry accuracy of the [OIII] frames was improved by using the
  corresponding V-band image plate solution after being registered to
  the V-band image coordinate system. On/off images need to be
  registered on the same system due to the selection of PNs via colour
  excess (see Sect.\ref{subsec:extr}).}.

We flux calibrated the broad-band and narrow-band frames to the AB
system by observing standard stars through the same filters used for
the survey. We used spectrophotometric standard star Hilt600 and
Landolt stars for the [OIII]-band and V-band flux calibrations
respectively, giving zero points for the [OIII] and V-band frames
equal to $Z_{\mathrm{\left[OIII\right]}}=24.29\pm0.04$ and
$Z_{\mathrm{\left[V\right]}}=27.40\pm0.17$, normalised to a 1 second
exposure\footnote{These values of $Z_{\mathrm{\left[OIII\right]}}$ and
  $Z_{\mathrm{\left[V\right]}}$ are consistent with those obtained by
  \citet{castro09} where they analyze data collected with the
  Suprime-Cam at the Subaru telescope through the same filters: [OIII]
  narrow-band filter ($ \lambda_{c}$=5029 \AA\, $\Delta\lambda$=74
  \AA), and broad-band V-filter ($ \lambda_{c}$=5500 \AA\,
  $\Delta\lambda$=956 \AA).}.

However, the integrated flux from the [OIII] line of a PN is usually
expressed by the magnitude $m_{5007}$ using the relation introduced by
\citet{jacoby89}:

\begin{equation}
  m_{5007}=-2.5\log_{10}F_{5007}-13.74,
\end{equation}

where $F_{5007}$ is in units of ergs cm$^{-2}$ s$^{-1}$. From this we
determined the absolute flux calibration for the nebular flux in the
[OIII] emission line, following \citet{jacoby87}.  This flux
calibration takes the filter transmission efficiency at the wavelength
of the emission line into account, due to the fact that the fast
optics of wide-field instruments, such as the Suprime-Cam at the
Subaru telescope can affect the transmission properties of the
interference filter, especially for narrow-band imaging.  This effect
was quantified making use of the filter transmission curve described
in \citet{arnaboldi03}, representing the expected transmission of the
[OIII] interference filter in the f/1.86 beam of the Suprime-Cam at
the Subaru telescope.

The relation between the $m_{5007}$ and $m_{\mathrm{AB}}$ magnitudes
for the narrow-band filter is given by:
\begin{equation}
  m_{5007}=m_{\mathrm{AB}}+2.49.
\end{equation}
A detailed description of the relation between AB and  $m_{5007}$[OIII]
magnitudes is given in \citet{arnaboldi02}.

In this paper we will use the notation of $m_{n}$ and $m_{b}$ to refer
to the narrow-band and broad band  magnitudes of the objects in the AB
system.  We  will use  $m_{5007}$  to  refer  to the  [OIII]  magnitudes
introduced by \citet{jacoby89}.

Table  \ref{table:2}  gives  the  constant value  for  AB-to-$m_{5007}$
magnitude  conversion,  and  limiting  magnitudes in  the  on-band  and
off-band frames analysed in this paper.
 
\begin{table}[!h]
  \caption{On-band and off-band limiting magnitudes.}
  \label{table:2}      
  \centering          
  \begin{tabular}{ c c c c} 
    \hline\hline\\

    Field & $m$$_{lim,5007}$ & $m_{lim,b}$& C \\
    \hline\\       
    \hline \\
                   
    M87 SUB1 &28.39 &26.4 &2.49\\
    M87 SUB2 &28.39 &26.6 &2.49
    \\
    \hline                  
  \end{tabular}

  \tablefoot{C is  the transformation constant between AB and
    5007  magnitudes  for  the   narrow-band  [OIII]  filter:  $m_{5007}  =
    m_{\mathrm{AB}} + $C.}
          
\end{table}

The final products for the M87~SUB1 and M87~SUB2 pointings consist of
the stacked images in the [OIII] and V band, astrometrically and
photometrically calibrated.  As consequence of the observing strategy
the seeing measured as the average FWHM of stellar sources is similar
in both pairs of images, and less than FWHM$_{\mathrm{seeing}}<1
\arcsec$ (see Table\ref{table:1}). A fit of the PSF with the IRAF task
{\bf psf} (in the digiphot/daophot package) shows that a Moffat
analytical function with a $\beta$ parameter $\beta=2.5$ (see
IRAF/digiphot manual for more details) is the best profile to model
the stellar light distribution of point sources in our
images. Moreover, the PSF fit was computed in three different regions
of the on-band and off-band images for both the M87~SUB1 and M87~SUB2
fields, covering the images from the centre to the edges: for all
regions examined, the PSF fit gave the same best fit solution, showing
that the PSF does not vary across the images.

 \section{Selection of PN Candidates and Catalogue Extraction}
\label{section3}

As a result of their bright [OIII] ($\lambda$5007) and faint continuum
emission, extragalactic PNs can be identified as objects detected in
images taken through the on-band [OIII] filter, but not detected in
images taken through the off-band continuum filter. In our work, all
PN candidates are extracted through the on-off band technique
\citep{jacoby90} using selection criteria based on the distribution of
the detected sources in colour magnitude diagrams \citep{thensus97}.
We used an automatic extraction procedure developed and validated in
\citet{arnaboldi02,arnaboldi03} for the identification of emission
line objects in our images.  We give a brief summary of the selection
procedure in the next section.

\subsection{Extraction  of   Point-Like  Emission-Line  Objects}
\label{subsec:extr} 
We employed the object detection algorithm SExtractor
\citep{bertin96}, that detects and measures flux from point-like and
extended objects.  Sources were detected in the on-band image
requiring that 20 adjacent pixels or more have flux values $1\times
\sigma$ RMS above the background.  Magnitudes were then measured in a
fixed aperture of radius $R= 6$ pixels (1.2$\arcsec$, $\sim3$ times the
seeing radius) for sources in the on-band image, and then through the
aperture photometry in the V-band image at the (x,y) positions of the
detected [OIII] sources with SExtractor in dual-image mode.
Candidates for which SExtractor could not detect a $m_{b}$ magnitude
at the position of the [OIII] emitter were assigned a $m_{b}=28.7$,
i.e.  the flux from an [OIII] emission of $m_{n}=m_{lim,n}$ seen
through a V-band filter \citep{thensus97}.  All objects were plotted
in a colour magnitude diagram (CMD), $m_{n}$-$m_{b}$ vs.  $m_{n}$, and
classified according to their positions in this diagram.  Based on
their strong [OIII] line emission, the most likely PN candidates are
point-like objects with colour excess corresponding to an observed EW
greater than 110 \AA, after convolution with photometric errors as
function of the magnitude.  The limit of completeness is defined as
the magnitude at which the recovery fraction of an input simulated
point-like population with a given luminosity function\footnote{In
  these simulations we adopted a Moffat $\beta=2.5$ profile to define
  the PSF of unresolved sources (see Sect.\ref{section2.2} for
  details) and an exponential LF similar to the one describing the
  PNLF in \citet{ciardullo98}.}, drops below a threshold set to
50\%. We find $m_{lim,5007}$=28.39; this is the limiting magnitude of
our sample.

The colour selection of PN candidates requires off-band images deep
enough so that the colour can be measured reliably at fainter
magnitudes. As for the [OIII] images, we define the V-band
limiting magnitude as the faintest magnitude at which half of the
input simulated sample is retrieved from the image; we then derive
$m_{lim,b}= 26.4$ and $m_{lim,b}= 26.6$ for the M87~SUB1 and M87~SUB2
fields respectively.

\subsubsection{Colour selection}\label{subcolsel}
PN candidates are defined as objects with [OIII] magnitudes brighter
than the [OIII] limiting magnitude and with a colour excess, $m_{n}
\le m_{lim,n}$ and $m_{n}-m_{b} < -0.99$, the latter representing the
colour excess corresponding to $\mathrm{EW}_{\mathrm{obs}}=110$
\AA. The relation between the observed EW and the colour in magnitudes
is given by \citet{teplitz00}:
$\mathrm{EW}_{\mathrm{obs}}\simeq\Delta\lambda_{\mathrm{nb}}(10^{0.4\Delta
  m}-1)$, where $\Delta\lambda_{\mathrm{nb}}$ is the width of the
narrow band filter and $\Delta m = m_b -m_n$ is the colour.  A value
of $\mathrm{EW}_{\mathrm{obs}}=110$ \AA\ limits contamination from
[OII]$\lambda$3726.9 emitters at redshift $z \sim 0.34$ (see Sect.
\ref{subsec:val}).

Photometric errors may be responsible for continuum emission sources
falling below the adopted colour excess, hence contaminating the
emission object distribution.  This effect is limited by defining
regions in which 99\% and 99.9\% of a simulated continuum population
would fall in this CMD. Below the 99.9\% line, the probability to
detect stars is reduced to the 0.1\% level.

Theoretically a PN is a point-like source with no detected continuum,
i.e.  with no broadband magnitude measured by Sextractor. Nevertheless
we expect a continuum contribution in an aperture at the position of a
[OIII] source in the halo of M87 because of crowding effects or
residuals from the subtraction of the continuum light from the M87
halo.  This is confirmed by the distribution of colours measured for a
simulated PN population in the outer regions of M87, which we
  discuss in Sect.\ref{submissPN}.

\subsubsection{Point-like  versus extended sources}\label{subpointlike}
\begin{figure} \centering
  \includegraphics[width=9cm]{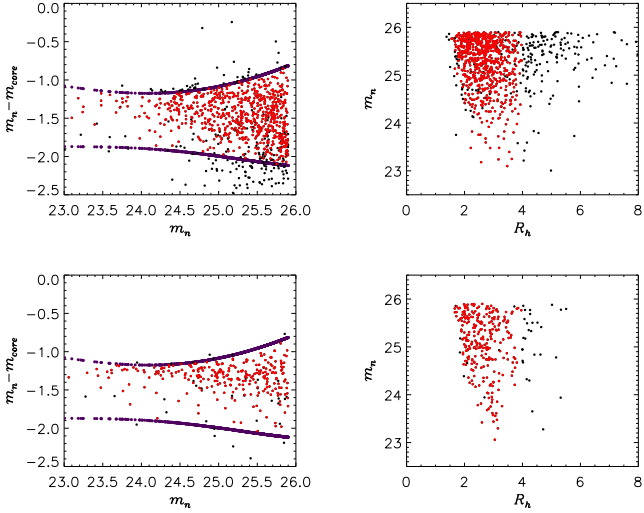}
  \caption{Point-source test, showing $m_{n}-m_{core}$ vs. $m_{n}$
    (left) and $m_{n}$ vs.  $R_h$ (right) for observed sources
    satisfying our colour restrictions in the M87 SUB1 field (top) and
    for modeled point-like emission objects (bottom).  The dark lines
    in the left panels delimit the region within which 96\% of the
    modeled point-like emission objects fall. Observed sources in the
    top panels are considered as point-like objects if they lie in
    this region and have a half light radius in the range $ 1 \le
    R_{h} \le 4$ pixels.  This range of $R_{h}$ was chosen such that
    95\% of the modeled population was included.  Red dots represent
    objects that satisfy both criteria and are thus selected as
    point-like objects. These criteria are also applied to sources in M87
    SUB2 field.}
  \label{ptest}
\end{figure}

At a distance of $\sim 15$ Mpc, PNs are unresolved points of green
light. We need to distinguish them from spatially extended background
galaxies and, possibily, HII regions. Point-like objects in our
catalogue are candidates that satisfy the following criteria:
\begin{enumerate}
\item we compare the SExtractor $m_{n}$ and $m_{core}$ magnitudes,
  where $m_{core}$ represents the measured magnitude in a fixed
  aperture of radius $R=2$ pixels (0.4$\arcsec$). For point-like
  objects $m_{n}-m_{core}$ has a constant value as function of
  magnitude, while it varies for extended sources.  We analyzed this
  difference for simulated point-like objects and determined the range
  of $m_{n}-m_{core}$ as function of $m_{n}$ where 96\% of the
  simulated population fall.  This range is narrow for bright
  magnitudes but it becomes wider towards fainter magnitudes due to the
  photometric errors.
\item we derive the distribution for the SExtractor half-light radius
  $R_{h}$, i.e.  the radius within which half of the object's total
  flux is contained.  For point-like objects this value should be
  constant but due to photometric errors it is confined to a range $ 1
  \le R_{h} \le 4$ pixels that we determine as the range in which 95\% of
  the simulated population lies.
\end{enumerate}
These criteria for the selection of point-like sources are shown in
Fig.~\ref{ptest}. For the simulated population, 91\% of the input
sources are recovered as point-like objects. Therefore, applying these
criteria to the observed objects excludes 9\% of the PNs from the
sample along with the extended objects.

%
\noindent

\subsubsection{Masking of bad/noisy regions}\label{submaskregion}

The last step of our automatic selection is the masking of the areas
where the detection and photometric measurements of sources are
dominated by non-Poisson noise. These areas include diffraction and
bleed spikes from the brightest stars, bad pixel regions and higher
background noise regions, the latter mostly at the edges of the images
because the dithering strategies lead to different exposure depth near
the edges (we will discuss this further in Section~\ref{subcatval}).
After these regions are excluded, the total effective area of our
survey is $\sim 0.43$ deg$^2$.

 The selection described above leads to a catalogue, hereafter called
 automatic sample, containing 792 objects classified as PN
 candidates. These sources are point-like objects with colour excess,
 corresponding to an observed equivalent width
 $\mathrm{EW}_{\mathrm{obs}}>110$ \AA\, and are located in regions of
 the colour-magnitude diagram where the contamination by foreground
 stars is below 0.1\%.

\begin{figure} \centering
  \includegraphics[width=9cm]{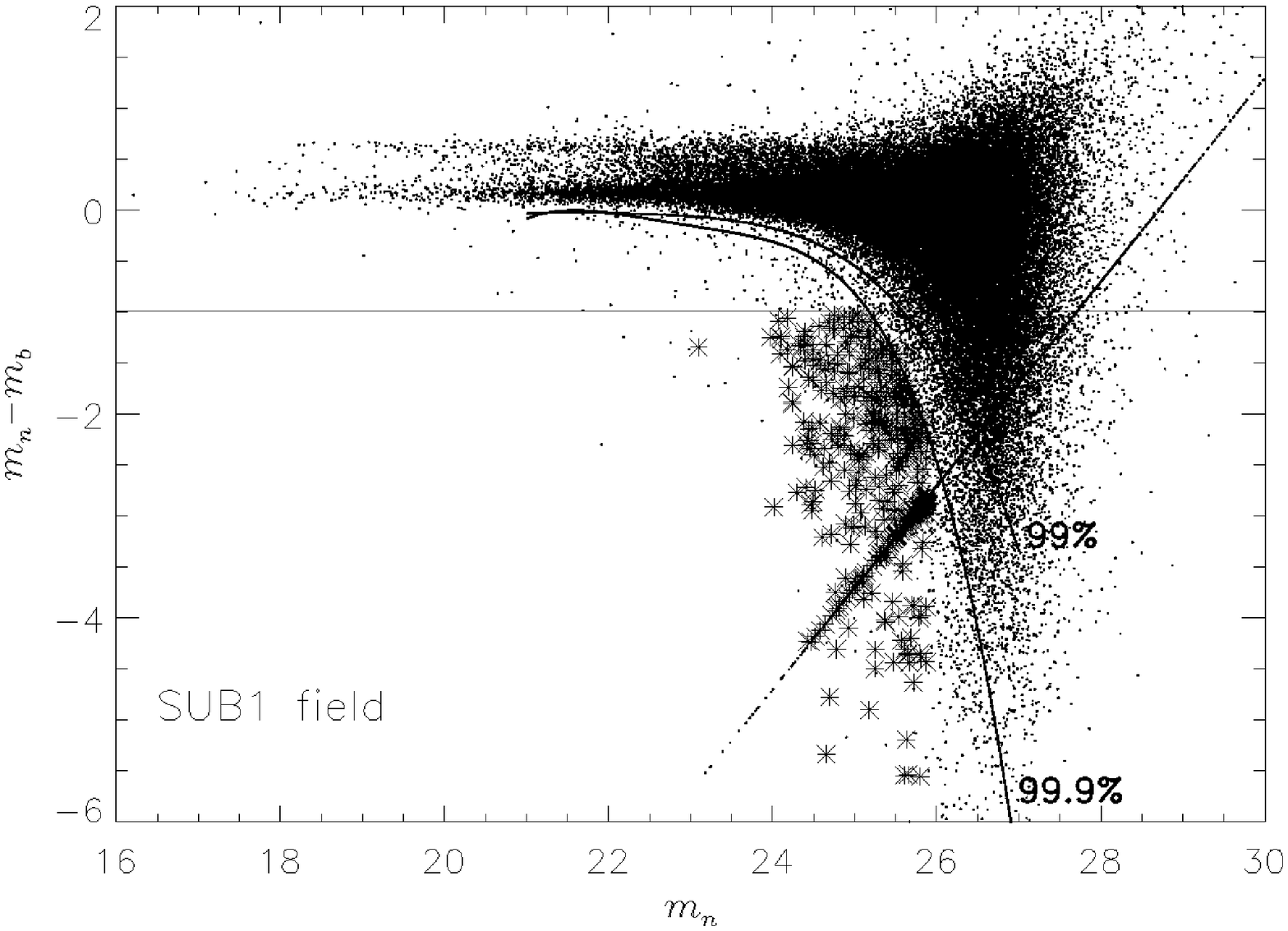}
  \includegraphics[width=9cm]{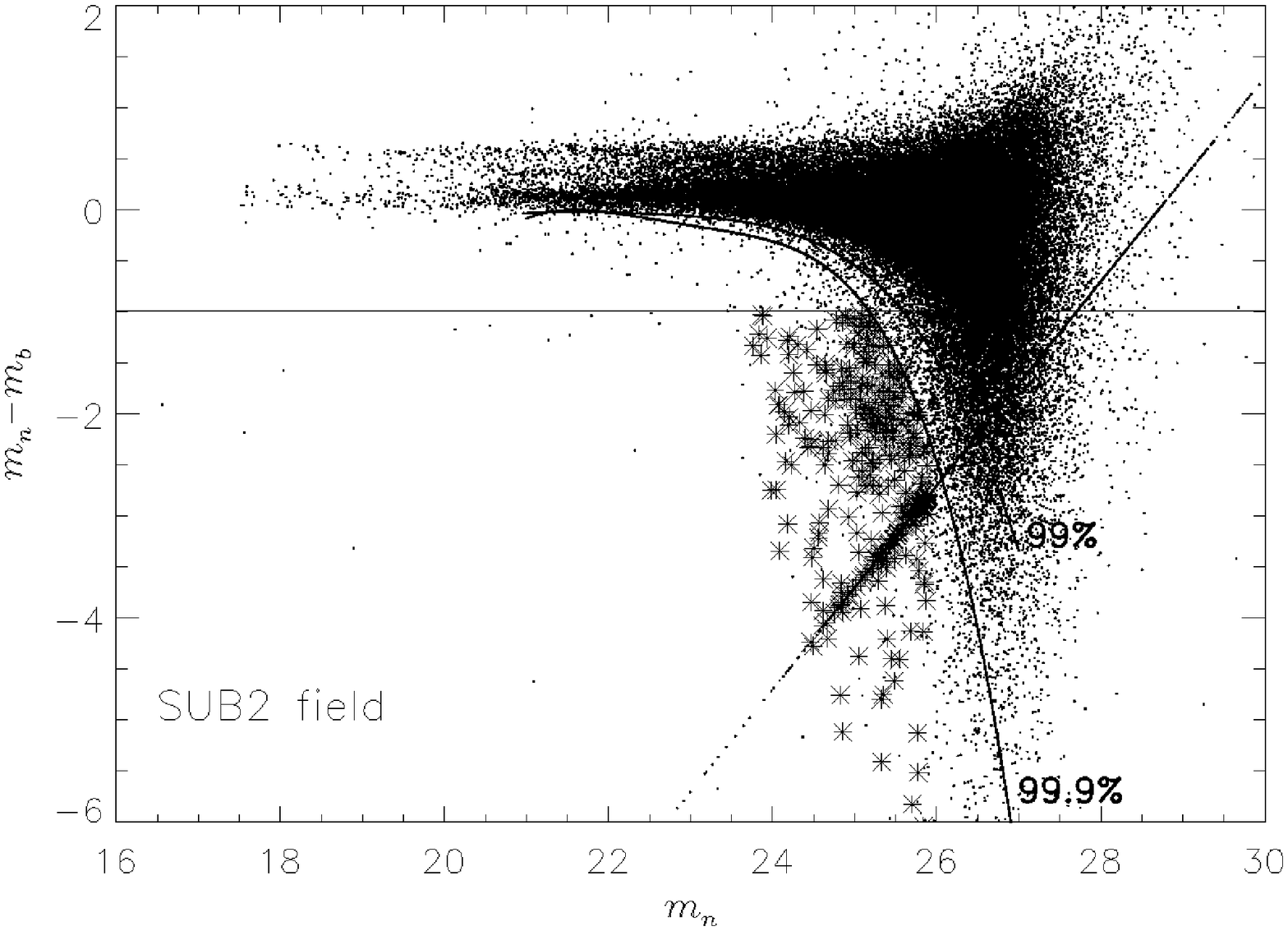}
  \caption{CMD for all sources in the M87~SUB1 (top panel) and
    M87~SUB2 (bottom panel) fields.  The horizontal lines indicate the
    colour excess of emission line objects with an
    EW$_{\mathrm{obs}}=110 $\AA. The curved lines delimit the regions
    above which 99\% and 99.9\% of the simulated continuum objects
    fall in this diagram, given the photometric errors.  The set of
    points on the inclined line represents those objects with no
    broadband magnitude measured by SExtractor (see
    Sect.~\ref{subsec:extr} for more details).  Asterisks represent
    objects classified as PNs according to the selection criteria
    discussed in Sect.~\ref{subsec:extr}.}
  \label{cmd}
\end{figure}

Fig.~\ref{cmd} shows the colour selection for the M87~SUB1 and M87~SUB2
fields, with the PN candidates represented by asterisks.

%

\subsubsection{Missing PNs in the photometric sample}\label{submissPN}

\begin{figure} \centering
  \includegraphics[width=9cm]{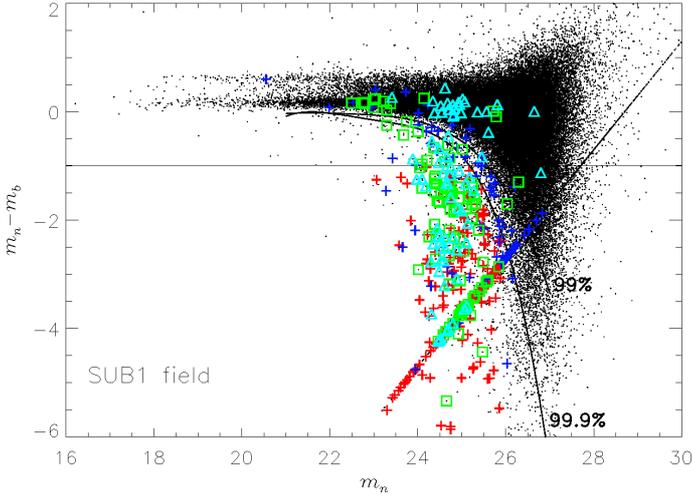}
  \caption{CMD for a simulated PN population in the M87~SUB1 field
    with intrinsic magnitudes $ 23 \le m_{n} \le m_{lim,n}$ (crosses).
    Because of photometric errors, our selection criteria would
    exclude 28.3\% of the input sample (blue crosses). The remaining
    71.7\% would be selected as PNs (red crosses). Dots surrounded by
    green squares and cyan triangles are emission objects in common
    between our sample and the PN samples selected by
    \citet{ciardullo98} and \citet{feldmeier03}, respectively. Solid
    lines as in Fig.~\ref{cmd}.}
  \label{f1_sim_CMD}
\end{figure}

The selection procedure based on flux thresholds is sensitive to
photometric errors. At fainter magnitudes, PN candidates with
intrinsic $\mathrm{EW}_{\mathrm{obs}} > 110$ \AA, or $m_{n} \le
m_{lim,n} $, can have smaller measured $\mathrm{EW}_{\mathrm{obs}}$,
or fainter $m_{n}$, because of photometric errors and are, therefore,
excluded from our selection.  We quantified this effect by simulating
an [OIII] emission line population, with an exponential LF in the
magnitude range $23 \le m \le 27.5$, randomly distributed on the
on-band scientific image.  No continuum emission was assigned to the
objects.  We then carried out the photometry as for the real sources:
their CMD is shown in Fig.~\ref{f1_sim_CMD}. We see that many
simulated PNs have a measured continuum magnitude due to crowding
effects or residuals from the galaxy background. This CMD shows also
that for simulated sources brighter than the limiting magnitude, the
percentage of PNs that we would miss due to photometric errors is
28.3\% and 29.8\% in the M87~SUB1 and M87~SUB2 fields
respectively. This implies that our automatic procedure, within the
magnitude range $ 23 \le m_{n} \le m_{lim,n}$, recovers 0.71 of the
total sample. In Tab.~\ref{table:PNLF} we show the dependence of this
\textit{colour completeness} on magnitude.

\subsubsection{Catalogue validation: visual inspection and joint
  candidates in both fields}\label{subcatval}

Finally, the photometric catalogue obtained from the automatic
selection described in the previous Sections was visually inspected,
in particular for regions near bright stars or with a higher noise level.
This visual inspection lead to a catalogue of 688 candidates, the removed
spurious detections being $\sim$11\% of the automatic extracted
sources.

In the final catalogue 18 candidates appear in both fields and their
magnitudes are measured independently in the two
pointings. Differences between the independent measurements are
consistent with the errors.

\subsection{Possible sources of contaminants in the PN sample}
\label{subsec:val} 
Following \citet{aguerri05}, we examined the main contaminants 
and estimate their contributions to the final catalogue.\\
 
\noindent
\emph{Contamination by faint continuum objects--} At
  $m_{lim,n}$, the faint continuum objects can mimic an [OIII]
  emission line population, because they are scattered into the region
  where PN candidates are selected. We constrained this contribution
  by computing the 99.9\% lines for the distribution of continuum
  objects.  From the total number of the observed foreground stars, we
  determine the number of objects (0.1\%) that would be scattered in
  the region of colour selected PNs. The resulting contribution from
  faint stars equals 9\% and 11\% of the total extracted sample for
  the M87~SUB1 and M87~SUB2 fields, respectively.

\noindent
\emph{Contamination by background galaxies: Lyman $\alpha$
  galaxies and [OII] emitters --} The strong [OIII] $\lambda$5007 PN
  emission with no associated detected continuum allows us to identify
  PNs as objects with negative colour \citep{thensus97}.  This colour
  selection will also identify Ly$\alpha$ galaxies at redshift
  $z\sim3.1$ as well as [OII] $\lambda$3727.26 emitters at redshift
  $z\sim0.34$, whose emission lines fall within the bandpass of our
  narrow band filter.

  The contamination by Ly$\alpha$ is quantified by using
  the number density of $z=3.1$ Ly$\alpha$ galaxies from
  \citet{gronwall07} where their limiting magnitude of
  $m_{lim,n(G07)}$(5007)=28.31 for statistical completeness makes
  their survey as deep as ours. We consider their Ly$\alpha$ LF
  given by a Schechter function of the form:
  
\begin{equation}
  \phi(L)d(L/L^*)\propto(L/L^*)^{\alpha}e^{-L/L^*}d(L/L^*),
\end{equation} 

with their best-fit values of $\mathrm{log_{10}}L^*=42.66$ erg
s$^{-1}$ and $\alpha=-1.36$, corrected for the effects of photometric
errors and their filter's non-square transmission curve. At redshift
$z=3.1$, the surveyed area of $0.43$ deg$^2$, observed through the
Suprime-Cam narrow-band filter, samples $\sim 3.1$ $10^{5}$ Mpc$^3$
\citep{hogg99}. However, as pointed out by \citet{gronwall07}, when
working with narrow band data taken through a nonsquare filter
bandpass, the effective survey volume is $\sim25\%$ smaller than that
inferred from the interference filter's FWHM. We thus compute the
number of expected Ly$\alpha$ emitters by using a survey volume of
$\sim2.3$ $10^{5}$ Mpc$^3$. The Ly$\alpha$ population at $z\sim3.1$
shows clustering over a correlation length of $r_{0}=3.6$ Mpc
\citep{gawiser07}, corresponding to an angular size in the Virgo
cluster of $r'_{0}\sim7.9\arcmin$. Hence we need to allow for the
large-scale cosmic variance in the average Ly$\alpha$ density, which
in our case is $\sim 20\%$ \citep{somerville04,gawiser07}.  From our
effective sampled volume, we predict, then, that the number of
expected Ly$\alpha$ contaminants is $25\% \pm 5\%$ of our total
catalogue.

The contribution from [OII] emitters is considerably reduced by
selecting candidates with observed equivalent width,
$\mathrm{EW}_{\mathrm{obs}}>110$ \AA, corresponding to a colour
threshold $m_{n}-m_{b}<-0.99$ \citep{teplitz00}.  This is because no
[OII] emitters with $\mathrm{EW}_{\mathrm{obs}}>95$ \AA\ have been
found \citep{colless90,hammer97,hogg98}. Note that in the
\citet{gronwall07} sample of Ly$\alpha$ galaxies at $z=3.1$, the
fraction of [OII] contaminants is considered to be negligible because
of their EW selection ($\Delta m_{(G07)} \sim-1$).  This means that if
there is a fraction of [OII] emitters with $EW_{\mathrm{obs}}>110$
\AA\ in our sample, then their contribution is accounted for by using
the Ly$\alpha$ LF from \citet{gronwall07}.

\subsection{Comparison with previous PN samples in M87}
\label{comparison_C98}

Our survey area overlaps with those studied by \citet{jacoby90},
\citet[hereafter C98]{ciardullo98} and \citet[hereafter
F03]{feldmeier03}; the locations of the C98 and F03 fields are
over-plotted on Fig.~\ref{mapMihos}.

We have a limited number of objects in common between \citet{jacoby90}
and our catalogue ($\sim 14 \%$ of their catalogue), because of the high
residual background in our images from the bright central regions of
M87. These fractions are larger for the C98 and F03 catalogues, and we
discuss them in turn.

C98 carried out an [OIII] $\lambda$5007 survey for PNs, covering a
$16\arcmin \times 16\arcmin$ field around M87. They identified 329 PNs
in the M87 halo, 187 of these in a statistically complete sample down
to $m_{5007}=27.15$.  Of the 329 sources selected by C98, 201 fall
within our surveyed area\footnote{This area does not include the
  regions affected by high background noise and bad pixels (see
  Sect.~\ref{submaskregion})}. Of these sources, 91\% are matched with
[OIII] detected objects in our survey, but only $\sim$ 60\% satisfy
our selection criteria for PN candidates, see the CMD
for the C98 candidates (green squares) in Fig.~\ref{f1_sim_CMD}.

\begin{figure} \centering
  \includegraphics[width=9cm]{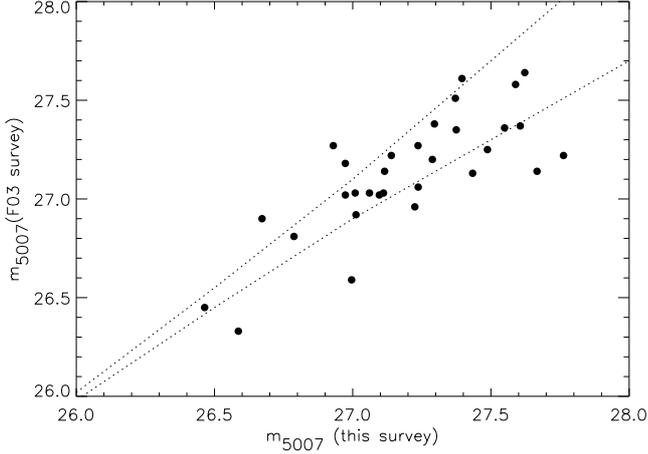} 
  \caption{$m_{5007}$ for the F03 PN candidates (corrected
      for $\sim 0.3$ mag offset) plotted against the $m_{5007}$
      magnitudes measured in our survey.  The dotted lines represent
      the 1$\sigma$ uncertainty from the photometric errors of our
      survey. The two magnitude systems are consistent within the
      photometric errors with an offset that is constant with
      magnitude.}
  \label{match}
\end{figure}

F03 carried out a survey of intracluster PNs (ICPNs) over several
fields in the Virgo cluster region; the one overlapping with the
current Subaru survey is a $16\arcmin \times 16\arcmin$ area north of
M87, labeled ``FCJ'' field in F03 and \citet{aguerri05}.  100\% of the
candidates in this field match with [OIII] sources in the M87 SUB1
field, but only 42\% satisfy the selection criteria for PN candidates
in our survey, see the CMD for the F03 candidates (cyan triangles) in
Fig.~\ref{f1_sim_CMD}.

On the basis of the common PN candidates with the largest S/N ratios,
we can compare the photometric calibration for the $m_{5007}$
magnitudes and any variations with magnitude in the different
samples. We find a constant offset that does not vary with magnitude
between the C98, F03 samples and our survey, with our $m_{5007}$
system being $\sim 0.3$ magnitudes fainter.  In Fig.\ref{match} the
F03 mags (corrected for $\sim 0.3$ mag shift) are compared with ours.

In Section~\ref{section5} we will compare the empirical PNLF for our
PN sample with those for the C98 and F03 data, and based on this
comparison will argue that there is a systematic effect in the C98 and
F03 photometry, causing brighter m$_{5007}$. In what follows, the
calibration offset is therefore applied to the previously published
data whenever we compare them with our PN magnitudes.

 \section{The radial profile of the PN population and comparison with
  the M87 surface brightness}
\label{section4}

In what follows we present the number density distribution from our PN
sample, one of the largest both in number of tracers and in radial
extent. Plotted in Fig.  \ref{spatial_dist} is the position of our PN
candidates together with the outline of our survey region.

\subsection{The PN radial density profile}
\label{density_p}

%
\begin{figure}
   \centering
   \includegraphics[width=9cm]{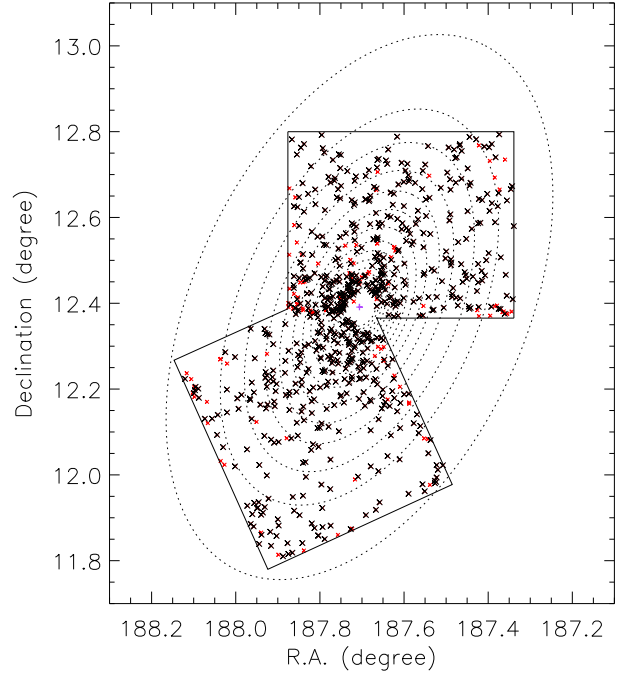}
   \caption{Spatial distribution of PN candidates (black crosses). Red
     crosses represent objects classified as spurious after visual
     inspection.  The magenta cross indicates the centre of M87.
     Dotted ellipses trace the M87 isophotes from $R=2.8$ to
     $R=40.7'$ along the photometric major axis, at a position angle
     P.A.=-25.6$^{\circ}$ \citep{kormendy09}.  The solid squares
     depict our survey area. North is up, East to the left.}
   \label{spatial_dist}
 \end{figure}

 We now investigate whether the PN number density profile follows the
 surface brightness profile of the galaxy light. On the basis of the
 simple stellar population theory, the luminosity-specific stellar
 death rate is insensitive to the population's age, initial
 mass function, and metallicity \citep{renzini86}. Thus the
 probability of finding a PN at any location in a galaxy should be
 proportional to the surface brightness of the galaxy at that
 location.

 In order to compare the surface brightness and PN number density
 profiles, we bin our PN sample in elliptical annuli, whose major axes
 are aligned with M87's photometric major axis and with ellipticities
 measured from the isophotes (see Fig.~\ref{spatial_dist}). We compute
 PN number densities as ratios between the number of PNs in each
 annulus and the area of the intersection of the annulus with our
 field of view. These areas, $A(R)$, are estimated
 using a Monte Carlo integration technique.

\begin{figure}
  \centering
  \includegraphics[width=9.5cm]{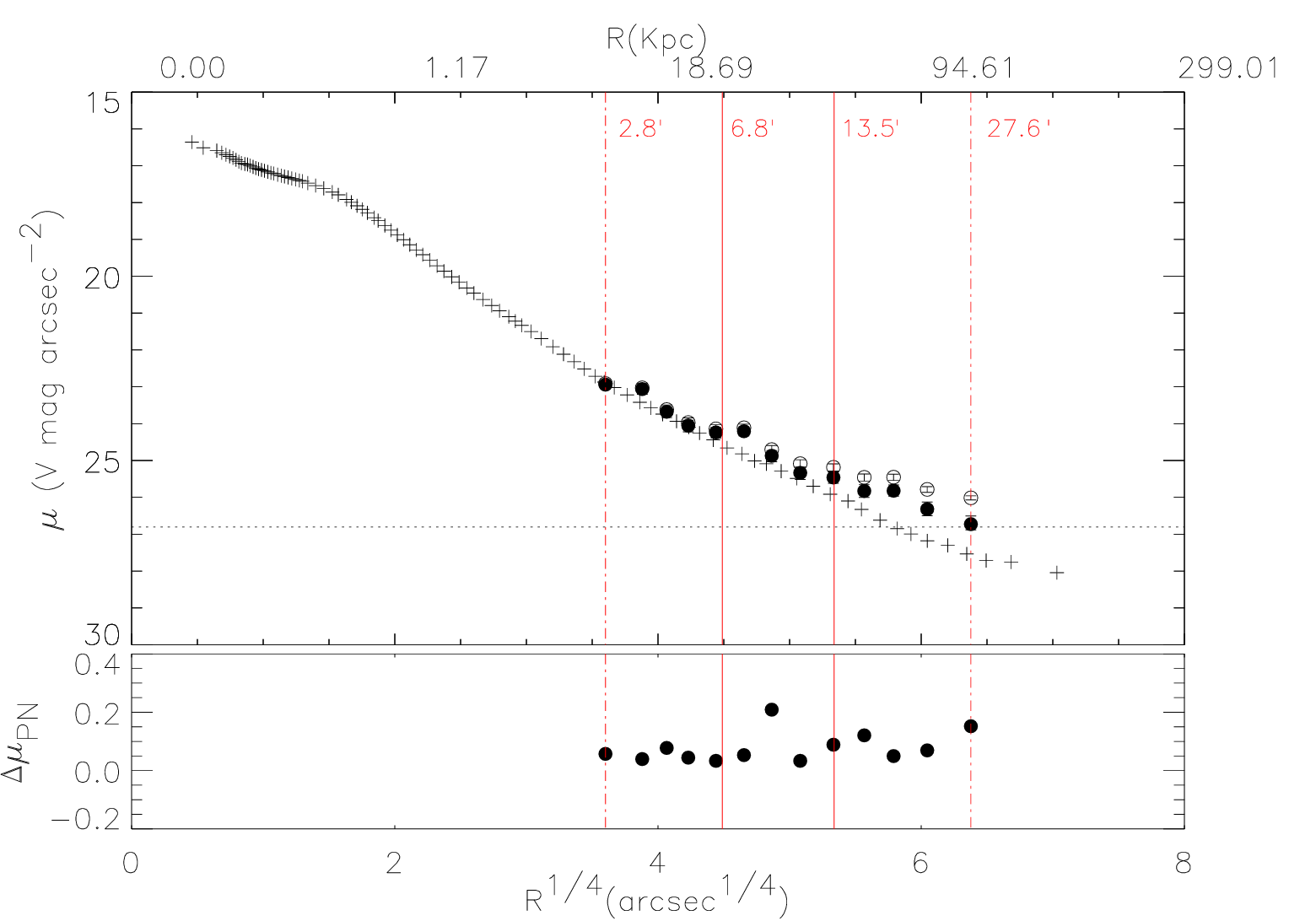}
  \caption{Top panel:~ comparison between surface brightness profile
    from \citet{kormendy09} (crosses) and logarithmic PN number
    density profile of the emission line candidates, brighter than
    $m_{5007}=28.3$, and corrected for spatial and colour
    incompleteness (open circles with error bars, see Table
    \ref{table:density_profile} for data), as function of the distance
    from the M87 centre. The black dotted line represents the
    contribution of Ly$\alpha$ emission objects to the logarithmic PN
    number density profile, assuming a homogeneous distribution over
    the surveyed area.  Under the same hypothesis, filled circles show
    the logarithmic PN number density profile when the Ly$\alpha$
    contribution is statistically subtracted.  Red lines mark the
    inner, intermediate and outermost regions of the M87 halo (see
    text). Bottom panel:~ difference between the logarithmic PN number
    density profiles if we had used the automatically extracted
    sources without the final inspection: a value of zero would mean
    that no variation in the number of sources is implied in the
    annulus at a given radius.}
  \label{density_profile}
\end{figure}

 The PN number density profile must be corrected for spatial
 incompleteness because the bright galaxy background and bright
 foreground stars may affect the detection of PNs. We compute this
 completeness function, C$(R)$, by adding a PN sample modelled
 according to a \citet{ciardullo89} PNLF on the scientific
 image. C$(R)$ is then the fraction of simulated objects recovered by
 Sextractor in the different elliptical annuli vs the input modelled
 population, for PNs brighter than $m_{5007}=28.3$ (values are given
 in Tab.\ref{table:density_profile}). The expected PN total number is
 then:
\begin{equation}
\mathrm{N_{\mathrm{c}}}(R)=\frac{\mathrm{N_{\mathrm{obs}}}(R)}{\mathrm{C}(R)0.71},
\end{equation}  
where the value 0.71 accounts for the average colour
incompleteness (see Sect.~\ref{submissPN}).  In
Fig.~\ref{density_profile} we show the comparison between the major
axis stellar surface brightness profile in the V band, $\mu_{\mathrm{K09}}$
\citep{kormendy09}, and the PN logarithmic number density profile, defined
as:

\begin{equation}
  \mu_{\mathrm{PN}}(R)=-2.5\log_{10}\left(\Sigma_{\mathrm{PN}}(R)\right)+\mu_0, 
  \label{rho}
\end{equation}
where
\begin{equation}
\Sigma_{\mathrm{PN}}(R)=\frac{\mathrm{N}_{\mathrm{c}}(R)}{A(R)}
\end{equation}
is the PN number density corrected for spatial and colour
incompleteness, and $\mu_0$ is a constant value added to match the PN
number density profile to the $\mu_{\mathrm{K09}}$ surface brightness
profile.  In the same plot, we also indicate radii that select three
different regions in the halo of M87; these regions are for $R <
\bar{R}/2$, $\bar{R}/2 \le R < \bar{R}$ and $R \ge \bar{R}$, where
$\bar{R}=13.5'$ represents the mean distance of the PN sample from the
centre of M87. The innermost and the outermost number density points
are at radii $R=2.8\arcmin$ and $R=27.6\arcmin$ respectively, and in
Fig.~\ref{density_profile} they are indicated with dotted red lines.
The surface brightness and the logarithmic PN number density profile
agree well in the innermost region, they slightly deviate in the
intermediate region while in the outermost region the logarithmic PN
number density profile flattens. The difference between the two
profiles amounts to 1.2 mag at the outermost radii.  This discrepancy
would still be observed if we had used instead the candidates from the
automatic selection procedure without any final inspection: the
difference between the two number densities is too small to affect
the slope at large radii (see Fig.~\ref{density_profile}, bottom
panel).  We also note that the flattening of the logarithmic number
density profile is seen in both Suprime-Cam fields independently.

Empirically, the logarithmic PN number density profile follows light
in elliptical \citep{coccato09} and S0 \citep{cortesi13}
galaxies. However, these studies cover the halos out to typically only 20
kpc.  The presence of Ly$\alpha$ background galaxies at
$z\sim3.1$ could also contribute to the flatter slope of the
logarithmic PN number density profile, and so we need to evaluate
their contribution to the number density at large radii. In
Fig.~\ref{density_profile} we show the contribution from Ly$\alpha$
contaminants (black dotted line) to the logarithmic number density
($25\% \pm 5\%$ of the total sample), assuming a homogeneous
distribution in the surveyed area. We can make this assumption because
the survey area $0.43$ deg$^{2}$ extends over many correlation lengths
of the Ly$\alpha$ population at $z=3.14$ (see Sect.~\ref{subsec:val}).

Hence we can statistically subtract the contribution of the Ly$\alpha$
backgrounds objects from the number of PN candidates in each annulus
and obtain the corrected number density profile (filled circles in
Fig.~\ref{density_profile}). Now the error bars also account for the
expected fluctuation of Ly$\alpha$ density (see Sect.~\ref{subsec:val}
for details) and the flattening of the logarithmic PN number density
profile is still observed. Using the derived $\alpha_{2.1}$ value for
the halo from Sect. \ref{phot_mod}, the surface brightness profile
translates to $\sim295$ PNs down to $m_{5007}=28.3$. Down to this
magnitude our catalogue contains $\sim$420 estimated PNs. The excess,
then, is $\sim 3.6 \times \sigma_{\mathrm{Ly}\alpha}$, where
$\sigma_{\mathrm{Ly}\alpha}$ is the standard deviation in the expected
number of Ly$\alpha$ emitters from cosmic variance and Poisson
statistics. Hence, the flattening can not be explained by fluctuations
in the Ly$\alpha$ fraction. In what follows we investigate the
physical origin of the flatter logarithmic PN number density profile.

%

\subsection{The $\alpha$ parameter}
The total number of PNs, N$_{\mathrm{PN}}$, is correlated with the
total bolometric luminosity of the parent stellar population,
$L_{\mathrm{bol}}$, through the so-called $\alpha$-parameter, that
defines the luminosity-specific PN density: N$_{\mathrm{PN}}=\alpha
L_{\mathrm{bol}}$.  

From stellar evolution theory it is found that the luminosity-specific
stellar death rate is insensitive to a stellar population's age,
initial mass function, and metallicity \citep{renzini86}. Therefore,
the total number of PNs associated with a parent stellar population
can be computed from the bolometric luminosity using the formula:
\begin{equation}
  \mathrm{N}_{\mathrm{PN}}=BL_{\mathrm{TOT}}\tau_{\mathrm{PN}},
  \label{N_PN}
\end{equation}
where $B$ is the specific evolutionary flux (stars yrs$^{-1}$
L$^{-1}_{\odot}$), $L_{\mathrm{TOT}}$ is the total bolometric
luminosity of the parent stellar population, and $\tau_{\mathrm{PN}}$
is the PN visibility lifetime. From Eq.~\ref{N_PN}, the
luminosity-specific PN number, $\alpha$, is
\begin{equation}
  \alpha=\frac{\mathrm{N}_{\mathrm{PN}}}{L_{\mathrm{TOT}}}=B\tau_{\mathrm{PN}}.
  \label{alpha}
\end{equation}
Observed values of the $\alpha$ parameter can then be interpreted as
different values of $\tau_{\mathrm{PN}}$ for the PNs associated with
different stellar populations, because variations of $B$ with
metallicity or Initial Mass Function (IMF) slope are
small\citep{renzini86}.

For any specific observation, the actual value of N$_{\mathrm{PN}}$
depends on the flux limit of the survey in which the PNs are
detected. Hence, for our survey depth we are interested in estimating
$\alpha_{2.1}$, the number of PNs within $\Delta m =2.1$ magnitudes of
the bright cutoff, per given amount of bolometric luminosity emitted
by the stellar population of the galaxy's halo or ICL.  For any
$\Delta m$, $\alpha_{\Delta m}$ is defined such that
\begin{equation}
\mathrm{N}_{\mathrm{PN,\Delta m}}=\int_{M^*}^{M^*+\Delta m}\! \mathrm{N}(m) \ \mathrm{d}m = \alpha_{ \Delta m}L_{\mathrm{bol}}.
\end{equation}
where N$(m)$ is the PN luminosity function and $M^*$ is the bright cutoff
magnitude. 

\subsection{Halo and Intracluster PN Population}
\label{alpha_par}
When studying the PN population in the outer region of M87 out to 150
kpc from the galaxy's centre, we expect contributions from the halo
PNs and from ICPNs. The presence of ICL in cluster cores is a
by-product of the mass assembly process of galaxy clusters
\citep{murante04,murante07}.

The existence of intracluster PNs in Virgo has been demonstrated on
the basis of extended imaging surveys in the 5007 \AA\ [OIII] line
\citep{feldmeier98,arnaboldi02,feldmeier03,aguerri05,castro09} and
spectroscopic follow-up \citep{arnaboldi96,arnaboldi03,arnaboldi04}.
From the spectroscopic follow-up of \citet{arnaboldi04} we know the
M87 halo and the Virgo core ICL to coexist for distances $R>16\arcmin$
from M87's centre (the FCJ field).  In fact, the projected phase space
diagram from \citet{doherty09} (PN line-of-sight velocity vs. radial
distance from M87 centre) shows the coexistence of the halo and ICL PN
population out to 150 kpc.

To estimate the ICL luminosity in our two fields, we assume a constant
surface brightness $\mu_{\mathrm{V}} =27.7$ mag arcsec$^{-2}$
\citep{mihos05}, and a V-band bolometric correction
BC$_{\mathrm{V}}$=-0.85 \citep{buzzoni06}, which then gives a total
bolometric luminosity in the ICL of $L_{\mathrm{ICL}}= 1.66 \times
10^{10} L_{\odot,\mathrm{bol}}$. This amounts to about one quoter of
the bolometric luminosity of the M87 halo.

\subsection{Two component photometric model for M87 halo
and ICL, and determination of their $\alpha_{2.5}$ parameters}
\label{phot_mod}
Therefore we now investigate whether the observed discrepancy between
the logarithmic PN number density profile and $\mu_{\mathrm{V}}$
surface brightness profile in Fig.\ref{density_profile} may be
explained by considering two PN populations associated with the M87
halo and ICL.  The physical parameter that links a PN population to
the luminosity of its parent stars is the luminosity-specific PN
number, the $\alpha$ parameter, thus a discrepancy between $\mu_V$ and
the measured logarithmic PN number density profile may come from
different $\alpha$ values for the halo and ICL stellar
population. This possibility is supported by \citet{doherty09} who
measured two different values of $\alpha$ for the bound (M87 halo) and
unbound (ICL) stellar component, which we are going to label
$\alpha_{\mathrm{halo}}$ and $\alpha_{\mathrm{ICL}}$ in what follows.
We can then define a photometric model with two components:

\begin{eqnarray}
\tilde{\Sigma}(R)  = \left[\alpha_{2.1,\mathrm{halo}}
  \mathrm{I}(R)_{\mathrm{halo,bol}}
  +\alpha_{2.1,\mathrm{ICL}}\mathrm{I}_{\mathrm{ICL,bol}}\right] 
\label{mod_sigma}\\
= \alpha_{2.1,\mathrm{halo}} \left[\mathrm{I}(R)_{\mathrm{K09,bol}}
    +\left(\frac{\alpha_{2.1,\mathrm{ICL}}}{\alpha_{2.1,\mathrm{halo}}}-1\right)\mathrm{I}_{\mathrm{ICL,bol}}\right] 
\label{mod_sigma_ratio}
\end{eqnarray}
where $\tilde{\Sigma}$(R) represents the predicted PN surface density
in units of $\mathrm{N}_\mathrm{{PN}}\mathrm{pc^{-2}}$,
$\mathrm{I}(R)_{\mathrm{halo}}$ and $\mathrm{I}_{\mathrm{ICL}}$ are
the surface brightnesses for the halo and the ICL components, and
$\mathrm{I}_{\mathrm{K09}}$ from \citet{kormendy09} is the observed
total surface brightness from M87 centre out to $40'$, accounting for
both halo and ICL components.  These surface brightnesses are in units
of $L_{\odot}\,\mathrm{pc^{-2}}$ and their bolometric values are
computed from the measured profiles in units of $\mathrm{mag}\
\mathrm{arcsec}^{-2}$ via the formula:
\begin{displaymath}
  \mathrm{I}=10^{-0.4\left(\mathrm{BC_V}-\mathrm{BC_{\odot}}\right)}10^{-0.4\left(\mu-\mathrm{K}\right)},
\end{displaymath}     
where BC$_{\mathrm{V}}$=-0.85 and BC$_{\odot}$=-0.07 are the V-band
and the Sun bolometric corrections, and $\mathrm{K}=26.4$
$\mathrm{mag}\ \mathrm{arcsec^{-2}}$ is a conversion factor from
$\mathrm{mag}\ \mathrm{arcsec^{-2}}$ to physical units
$L_{\odot}\mathrm{pc^{-2}}$ in the V-band. Assuming a fixed value of
BC$_{\mathrm{V}}=-0.85$ for every galaxy type has a 10 per cent
internal accuracy, i.e. $\pm$ 0.1 mag, for a range of simple stellar
population (SSP) models ranging from irregular to elliptical galaxies
\citep{buzzoni06}.

The surface brightness profile can be expressed in terms of PN surface
density $\tilde{\Sigma}(R)$:
\begin{equation}
  \tilde{\mu}(R)=-2.5\log_{10}\tilde{\Sigma}(R)+\mu_0,
  \label{mod_mu}
\end{equation}
where $\mu_0$ is a function of the $\alpha_{2.1,\mathrm{halo}}$ parameter:
\begin{equation}
  \mu_0=2.5\log_{10}\alpha_{2.1,\mathrm{halo}}+\mathrm{K}+(\mathrm{BC_{\odot}-BC_V}).
\label{mu_0}
\end{equation}

The value of $\mu_0$ is given by the value of the constant offset used
in Eq.~\ref{rho}, and it can be fixed by fitting this offset between
the observed logarithmic PN surface density and the surface brightness
profile, $\mu_V$, at smaller radii ($R \le 6.8\arcmin$).  From the
fitted offset of $\mu_0= 16.0 \pm 0.1$ mag arcsec$^{-2}$, we compute
the value for $\alpha_{\mathrm{2.1,halo}}$ (Eq.~\ref{mu_0})
resulting in $\alpha_{\mathrm{2.1,halo}}= (0.63 \pm 0.08)
\times 10^{-8}$ PN~L$_{\odot,\mathrm{bol}}^{-1}$ . The error on the
determined $\alpha_{\mathrm{2.1,halo}}$ is computed from the
propagation of the errors on the variables $\mu_0$ and
$\mathrm{BC_V}$, the latter having a 10\% accuracy.

\begin{figure}
  \centering
  \includegraphics[width=9cm]{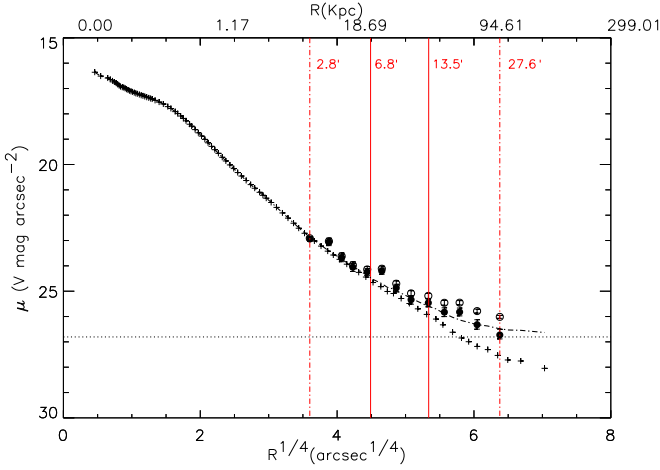}
  \caption{As in the top panel of Fig. \ref{density_profile} but with
    the modelled surface brightness profile from the two component
    photometric model superposed (dot-dashed black line).
    This modelled profile reproduces the flattening observed in
    the logarithmic number density well.}
  \label{mod_density_profile}
\end{figure}

In Fig.~\ref{mod_density_profile} we show the fit of the two component
PN model to the observed PN logarithmic number density profile for
$\alpha_{2.1,\mathrm{ICL}}/\alpha_{2.1,\mathrm{halo}}=3$ and a
constant ICL surface brightness of $\mu_{\mathrm{ICL}}=27.7$
$\mathrm{mag}\ \mathrm{arcsec^{-2}}$ \citep{mihos05}. The proposed
model predicts a flatter slope for the PN logarithmic number density
profile than the slope of the V-band surface brightness profile
$\mu_\mathrm{V}$, as observed. The fitted value for
$\alpha_{\mathrm{2.1,ICL}}$ is then $\alpha_{\mathrm{2.1,ICL}}= (1.89
\pm 0.29) \times 10^{-8}$ PN~L$_{\odot,\mathrm{bol}}^{-1}$.

Eq.~\ref{mod_sigma_ratio} shows explicitly that when
$\alpha_{2.1,\mathrm{halo}}=\alpha_{2.1,\mathrm{ICL}}$ then the
logarithmic PN number density profile should closely follow light.

The measured values of $\alpha_{\mathrm{2.1,halo}}$ and
$\alpha_{\mathrm{2.1,ICL}}$ correspond, down to the survey
depth, to an estimated $\sim390$ halo PNs, and $\sim310$ ICPNs in the
completeness-corrected sample, and to $\sim230$ halo PNs and $\sim190$
ICPNs in the observed sample.

To compute $\alpha_{2.5}$ for a PN sample which is not complete to
$m^*+2.5$, but only to a magnitude $m_{c}<m^*+2.5$, then
$\alpha_{2.5}$ is extrapolated by

\begin{equation}
\alpha_{2.5} = \Delta_{m_{c}}\times\alpha_{m_{c}}
\label{alpha_delta}
\end{equation}

where 

\begin{equation}
  \Delta_{m_{c}}=\frac{\int_{M^*}^{M^*+2.5}\! \mathrm{N}(m) \ \mathrm{d}m}{\int_{m^*}^{m_{c}}\! \mathrm{N}(m) \ \mathrm{d}m},
\label{delta}
\end{equation}

If we use the derived M87 PNLF (see Sect.~\ref{section5}) to compute
the cumulative number of PNs expected within 2.5 mag from the bright
cut off, normalised to the cumulative number in the first 2.1
magnitudes, we obtain $\Delta_{2.1}\simeq1.7$. As a result, the predicted
values of $\alpha_{2.5}$, for the halo and ICL components are
$\alpha_{2.5,\mathrm{halo}}=(1.10^{+0.17}_{-0.21})\times 10^{-8}$
N$_{\mathrm{PN}}$L$^{-1}_{\odot,\mathrm{bol}}$ and $
\alpha_{2.5,\mathrm{ICL}}=(3.29^{+0.60}_{-0.72})\times10^{-8}$
N$_{\mathrm{PN}}$L$^{-1}_{\odot,\mathrm{bol}}$ . The errors on the
$\alpha_{2.5}$ values also take the uncertainty on the
magnitude of the bright cutoff into account.
%

\subsection{Comparison with previously determined $\alpha_{2.5}$
  values}\label{DoDu}
The $\alpha_{2.5}$ values for the M87 halo and ICL were measured in
previous works by \citet{durrell02} and \citet{doherty09}. There are
several assumptions that are made when computing these values, hence
is important to address them before the actual numbers are compared.
As already pointed out in the previous section, when $\alpha_{2.5}$ is
computed for a PN sample which is not complete to $m^*+2.5$, then
$\alpha_{2.5}$ is extrapolated by following Eq.\ref{alpha_delta}.

In the studies of \citet{durrell02} and \citet{doherty09}, the PN
samples were complete one magnitude down the bright cutoff of
the PNLF. Then, $\alpha_{2.5}$ was computed using the extrapolation on
the basis of the analytic formula for $\mathrm{N}(m)$ by
\citet{ciardullo89}.

In Sect.~\ref{section5} we derive the M87 PNLF in the brightest 2.5
mag range, which shows a steeper slope at $\sim$ 1.5 mag below the
cutoff than what is predicted by the analytical formula. If we use
the observed PNLF then the cumulative number of PNs expected within
2.5 mag from the bright cut off, normalised to the cumulative number
in the first magnitude, is about 2.7 times that predicted by the
analytical PNLF. When comparing the actual values for the
$\alpha_{2.5}$ measured by \citet{durrell02} and \citet{doherty09},
they need to be rescaled by this fraction.

In the current work and in \citet{doherty09}, $\alpha_{2.5}$ values
were measured for the halo and ICL separately, using only the PNs
associated with each component. In \citet{doherty09} the line-of-sight
velocity of each PN was used to tag the PN candidate as M87 halo or
ICL. In \citet{durrell02} this information was not
available. Following \citet{doherty09}, only 58\% of the PN candidates
considered by \citet{durrell02} are truly ICL. When this correction is
applied, then the value measured by \citet{durrell02} for the ICPNs is
$\alpha_{2.5,\mathrm{ICL}}=1.3\times 10^{-8}$
PN~L$_{\odot,\mathrm{bol}}^{-1}$. If in addition we correct this
$\alpha_{2.5}$ value for the ICL for the different shape of the PNLF,
we get $\alpha_{2.5,\mathrm{ICL,DU02}}=3.6\times 10^{-8}$
PN~L$_{\odot,\mathrm{bol}}^{-1}$. This value is $\sim10\%$ greater
than ours, but consistent within the uncertainties and the
contamination from continuum sources in the F03 catalogue (see
Sect.\ref{comparison_C98}).

When we correct the $\alpha_{2.5}$ values for halo and ICL from
\citet{doherty09} by a factor 2.7, we obtain
$\alpha_{2.5,\mathrm{halo,D09}}=8.4\times 10^{-9}$
PN~L$_{\odot,\mathrm{bol}}^{-1}$ and
$\alpha_{2.5,\mathrm{ICL,D09}}=1.9\times 10^{-8}$
PN~L$_{\odot,\mathrm{bol}}^{-1}$. These values are smaller than ours,
but consistent within the uncertainties and the colour/detection
completeness corrections made here (see Sect.~\ref{density_p}) but not
in \citet{doherty09}.

\subsection{Implications of the measured $\alpha_{2.5}$ values}
According to the analytical formula of \citet{ciardullo89} for the
PNLF, $\alpha_{2.5}$ equals $\sim1/10$ of the total
luminosity-specific PN number $\alpha$ (see also \citet{buzzoni06}),
assuming that PNs are visible down to 8 mags from the bright cutoff
\citep{ciardullo89}. If we use this analytical function to extrapolate
the total number of PNs from 2.5 to 8 magnitudes below the bright
cutoff our measured values for $\alpha_{2.5}$ translate to total
luminosity-specific PN numbers for the two components of
$\alpha_{\mathrm{halo}} = 1.1 \times 10^{-7}$
PN~$L_{\odot,\mathrm{bol}}^{-1}$ and
$\alpha_{\mathrm{ICL}}=3.3 \times 10^{-7}$
PN~$L_{\odot,\mathrm{bol}}^{-1}$. We can recast these numbers in terms
of the PN visibility lifetime $\tau_{PN}$ through Eq.\ref{alpha},
assuming $B=1.8 \times 10^{-11} L^{-1}_{\odot}yr^{-1}$
\citep{buzzoni06}. $\tau_{PN}$ is then $\simeq 6.1 \times 10^{3}$ yr
and $\simeq 18.3 \times 10^{3}$ yr for the M87 halo and ICL PNs. Note,
that these would be lower limits, because at this stage we do not know
whether the extrapolation from 2.5 magnitudes below the cutoff to
fainter magnitudes follows the analytical formula, or whether it is
steeper.

Observationally there is some evidence that $\alpha$ is on average
larger for bluer systems \citep{peimbert90,hui93}. The measurements of
the colour profile in M87 show a bluer gradient towards larger radii
\citep{liu05,rudick10}: a fit of the colour profile along the major
axis inside 1000$\arcsec$ has a slope of -0.11 in
$\Delta(B-V)/\Delta\log(R)$ \citep{rudick10}.  The observed PN
logarithmic number density profile measured in this work is consistent
with the empirical result of a gradient towards bluer colours at
large radii. In the proposed model, the gradient is caused by the
increased contribution of ICL at large radii, which is bluer than the
M87 halo population.

One can ask whether the metallicity of the parent stellar population
may influence the number of PNs associated with a given bolometric
luminosity.  We are interested in population effects in the advanced
evolutionary phases of stellar evolution, as the [OIII] 5007
\AA\ emission and its line width are weakly dependent on the chemical
composition of the nebula \citep{dopita92,schonb10}.  For stellar
populations with the same IMF and ages, \cite{weiss09} computed models
for the asymptotic giant branch (AGB) for stars between 1.0 and 6.0
$M_\odot$ with different metallicities. Their results show that the
number of AGB stars varies with the chemical composition, because the
latter effects the lifetime on the thermally pulsing AGB, with the
longest lifetime obtained for metallicity between half and one tenth
of $Z_\odot$. Given that the evolutionary path followed by a star from
the end of the AGB to the beginning of the cooling phase of the
central white dwarf corresponds to the planetary nebulae phase, it is
suggestive to infer that stellar populations with metallicity -0.5 to
-0.1 solar may have a larger number of PNs than stellar populations
with solar metallicity or higher, for the same bolometric luminosity.
The metallicities of ICL stars in the Virgo cluster core were measured
by \citet{williams07} using colour magnitude diagrams to be between
-0.5 to -0.1 solar.  If the stars in the M87 halo have a higher
metallicity, we might expect a variation of the luminosity specific PN
number in the region of radii where the M87 stellar halo and the ICL
are superposed along the line-of-sight \citep{doherty09}, as is
quantified for the first time in this work.

 \section{Planetary Nebula Luminosity Function in the outer regions of M87}
\label{section5}

\subsection{PNLF}
\begin{figure}
  \centering
  \includegraphics[width=9cm]{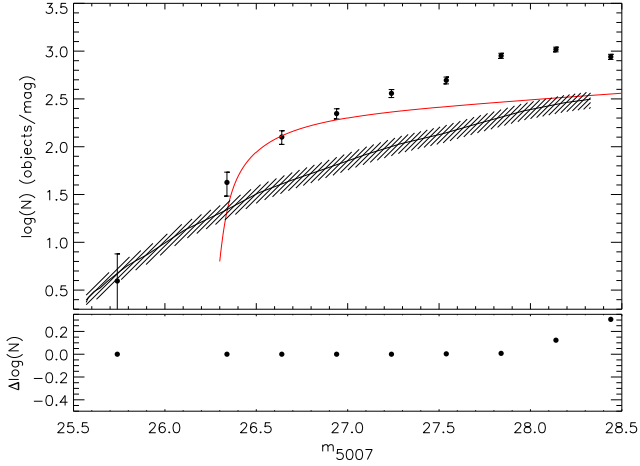}
  \caption{Top panel:~ luminosity function of the selected PNs in the
    outer regions of M87, corrected for colour and detection
    incompleteness. The error bars show the 1 $\sigma$ uncertainty
    from counting statistics.  Data are binned into 0.3 mag intervals
    (see Table~\ref {table:PNLF} for numerical values).  The solid red
    line represents the analytical PNLF model for a distance modulus
    of 30.8, convolved with photometric errors. The black line shows
    the Ly$\alpha$ LF from \citet{gronwall07}, scaled to the effective
    surveyed volume of the M87~SUB1 and M87~SUB2 fields, with the
    shaded area showing the cosmic variance due to Ly$\alpha$ density
    fluctuations ($\sim 20\%$, see Sect.~\ref{subsec:val}).  Bottom
    panel:~ difference between the PNLF if we had used instead the
    automatically extracted sources without any final inspection. This
    plot shows that spurious detections affect mostly the two faintest
    magnitude bins.}\label{PNLF}
\end{figure}

We can now use our large PN sample in the outer regions of M87 to
investigate the properties of the PNLF. To do this we need to take
into account that the fraction of detected PNs on our scientific
images can be affected by incompleteness, and that this incompleteness
is a function of magnitude. As for the spatial completeness, we
quantify this effect by adding a modelled sample of PNs on the
scientific images and computing the fraction recovered by Sextractor
of this input simulated PN population as function of magnitude (see
Tab.~\ref{table:PNLF} for values). In Fig.~\ref{PNLF} we show the PNLF
for the extracted candidates corrected for colour and detection
incompleteness. We can compare this PNLF with the analytical formula
by \citet{ciardullo89}
\begin{equation}
  \mathrm{N}(M) = c_1e^{c_{2}M}\left\{1-e^{3(M^*-M)}\right\}
  \label{PNLF_eq}
\end{equation}
where $c_1$ is a normalisation constant, $c_{2}=0.307$ and
$M^*$(5007)=-4.51 mag is the absolute magnitude of the PNLF bright
cutoff \citep{ciardullo89}. In Fig.~\ref{PNLF} the red solid line
shows the prediction from the analytical formula for a distance
modulus of 30.8, after convolution with photometric errors and
normalisation to the brightest observed bins.
%
The comparison of the derived PNLF with the analytical formula shows
two differences. First we detect one object with a $m_{5007}=25.7$
that is $\sim$0.6 mag brighter than the expected cutoff for a
distance modulus of 30.8. Second we measure a slope in the PNLF at
$\sim 1.5$ mag below the cutoff that is steeper than what is
predicted by the analytical formula in this magnitude range
\citep{ciardullo89,henize63}.

We discuss the significance of these deviations in turn.

{\it Over-luminous source --} Fig.~\ref{PNLF} also shows the
comparison of the PNLF from our PN sample with the Ly$\alpha$ LF from
\citet{gronwall07}, scaled to our effective surveyed volume. The
hatched range gives the uncertainty in the latter. We see that the
bright end of the Ly$\alpha$ LF is consistent with the luminosity of
the over-luminous object, within the photometric errors. Earlier
suggestions that such overluminous objects could be due to a depth
effect from Virgo ICL \citep[][C98]{jacoby90} are not consistent with
more recent studies of the ICL in Virgo; see
Section~\ref{PNLFoffset}. To definitively resolve the question of the
nature of this object, whether it is a Ly$\alpha$ emitter or an object
in the M87 halo, requires spectroscopic follow-up.

\begin{figure}
  \centering
\includegraphics[width=9cm]{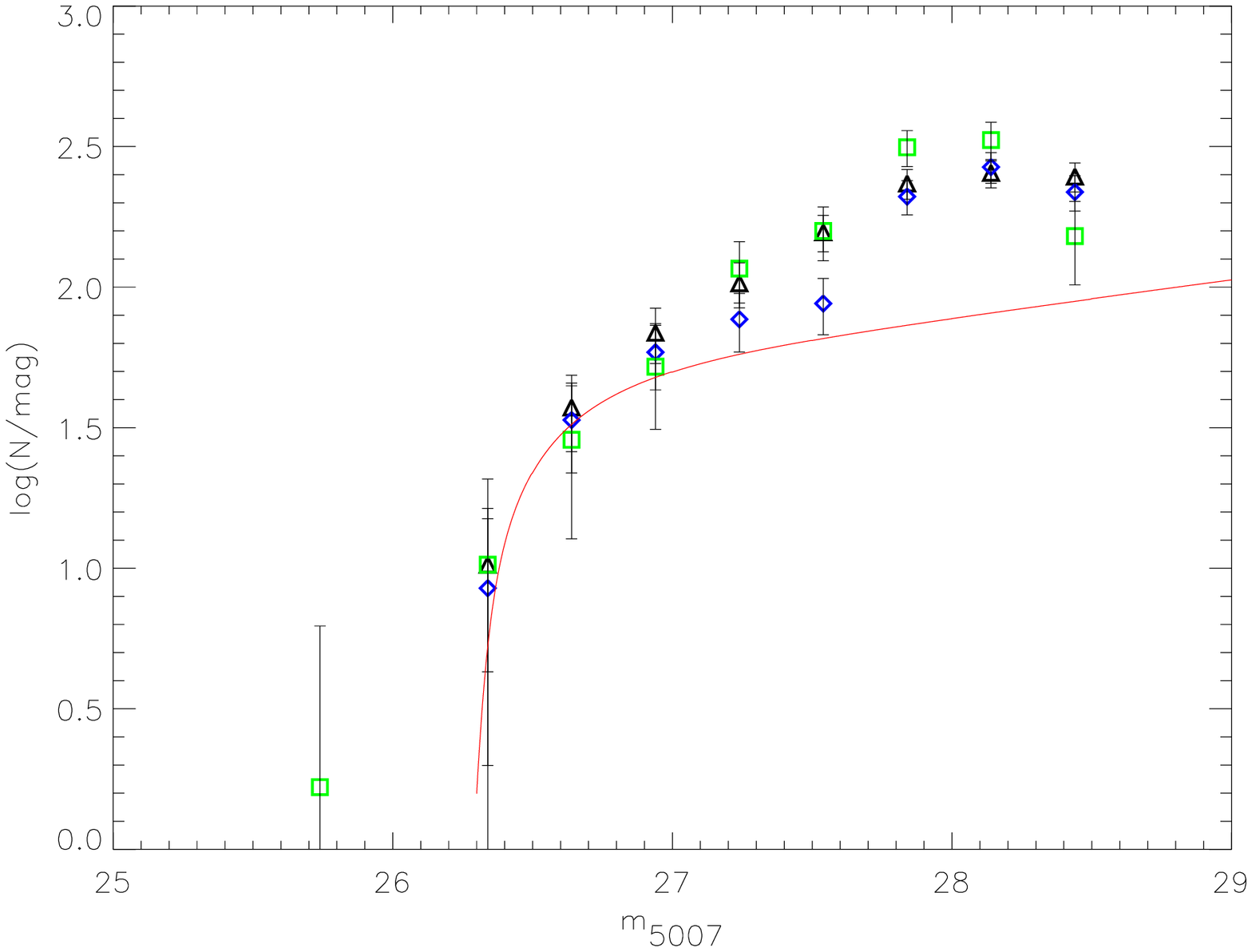}  
\includegraphics[width=9cm]{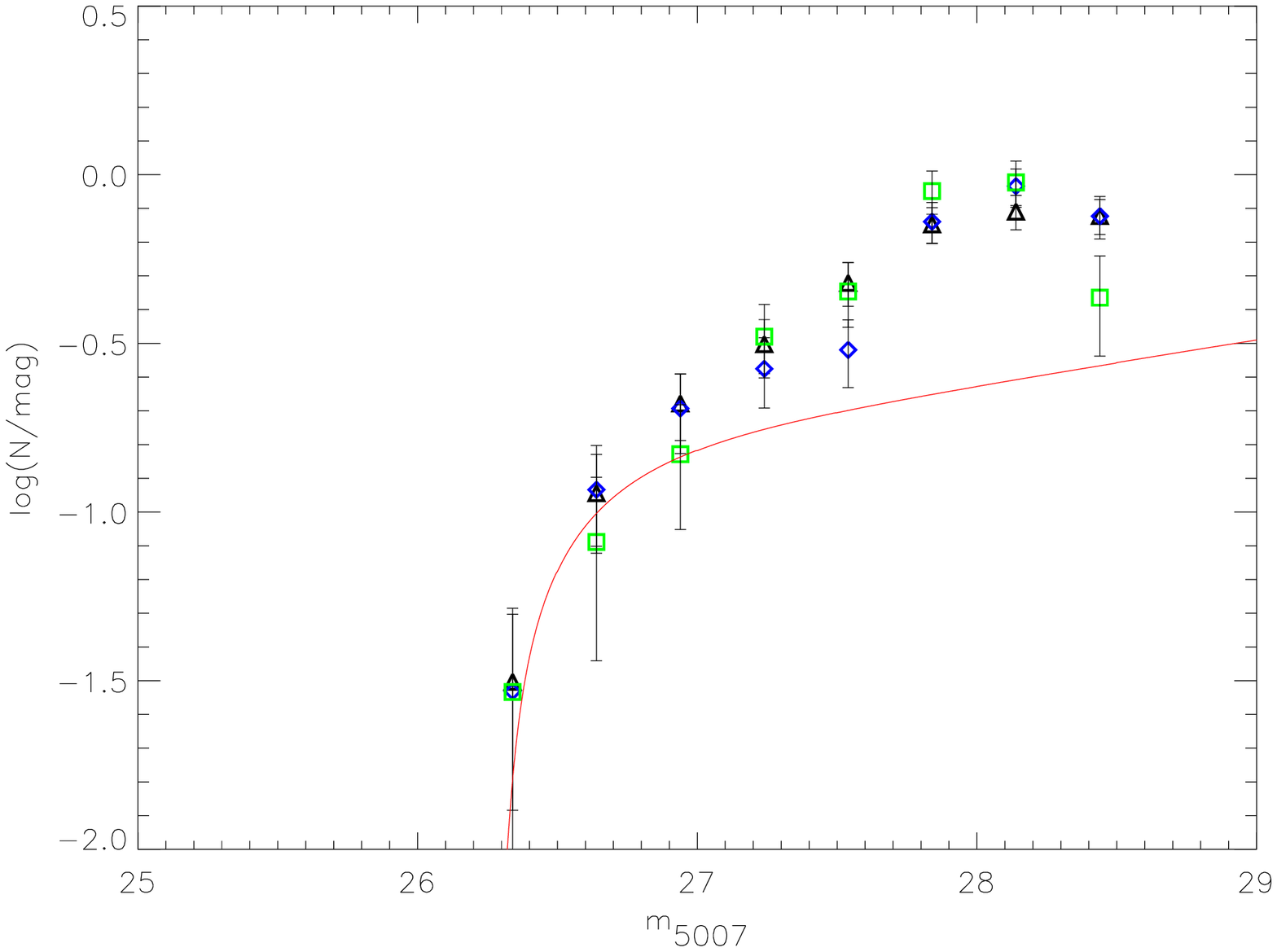}  
\caption{Top panel:~ empirical PNLFs in three radial ranges corrected
  for colour and detection incompleteness: PN candidates within
  6.5$\arcmin$ from M87 centre - triangles, PN candidates between
  6.5$\arcmin$ and 13.5$\arcmin$ from M87 centre - diamonds, PN
  candidates in the outermost region (distances greater than
  13$\arcmin$) - squares.The respective Ly$\alpha$ contribution
  expected in each radial bin was subtracted. Magnitudes are binned in
  0.3 mag bins and the error bars represent the 1 $\sigma$ uncertainty
  from counting statistics combined with the uncertainty from cosmic
  variance in the Ly$\alpha$ density (see text for details). As
  before, the red solid line is the convolved analytical formula of
  \citet{ciardullo89} for distance modulus 30.8.  Lower panel:~ same
  as for the upper plot, but now the three PNLFs are normalised at the
  total number of objects in each radial bin. The three data sets are
  consistent with being drawn from the same underlying distribution.}
  \label{PNLF_R}
\end{figure}

{\it Steeper PNLF 1.5 mag below the bright cutoff --} First, we
carried out a Kolmogorov-Smirnov test to check whether our empirical
PNLF can be drawn from the \citet{ciardullo89} analytical function,
and this possibility is rejected.  Next, because our sample covers
$0.5^\circ$ in the M87 halo, we can investigate the radial variation
of the PNLF out to radii of $30\arcmin$. We compute the PNLF for each
of the three PN subsamples associated with the radial bins described
in Sect.~\ref{density_p}, correcting for colour and detection
incompleteness, and subtracting the respective contribution from
Ly$\alpha$ background objects. These are shown in the upper panel of
Fig.~\ref{PNLF_R}, where the error bars also account for $\sim 20\%$
cosmic variance in the Ly$\alpha$ density (see
Sect. \ref{subsec:val}). We then compare the three PNLFs after
normalisation to the total number of objects in each radial bin and
carry out a Kolmogorov-Smirnov test to check whether they can result
from the same underlying distribution. The probability that these
three PNLFs are extracted from the same distribution is high, $P_{KS}
> 99\%$.

These results are significant because the steepening of the PNLF is
thus shown to be present in all three radial bins, while we know that
the ICPN population contributes mostly to the outermost bin, see
discussion in Sect.~\ref{alpha_par}, and that any residual
contamination of Ly$\alpha$ background emitters is also largest in the
outermost radial region. We also checked that the PNLFs in the two
Suprime-Cam fields are similar; both show the steepening at faint
magnitudes.  Hence we must conclude that the observed steepening of
the PNLF is an intrinsic property of the PN population (halo and ICL)
in the outer regions of M87.

\subsection{Generalised analytical model for the PNLF and distance
  modulus of M87 halo}
\label{distmod}

\begin{figure}
  \centering
\includegraphics[width=9cm]{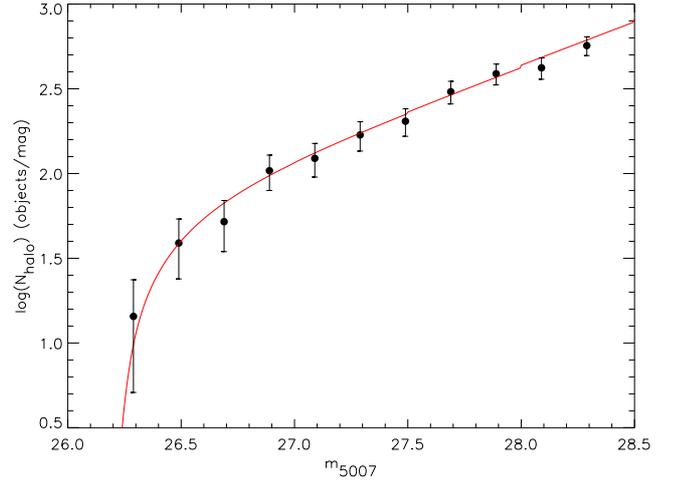}  
\caption{Completeness-corrected PNLF for the M87 halo (full
    dots) with magnitudes binned in 0.2 mag bins; error bars include
    Poisson statistics and 20\% variance in the number of subtracted
    Ly$\alpha$ contaminants. The continuous line indicate the resulting
    fit of the generalised analytical formula Eq.~\ref{PNLF_eq}, for
    the values of the free parameters given by $c_1=2017.2$,
    $c_2=1.17$ and a distance modulus of $m-M = 30.74$. The
    generalised PNLF is convolved with the photometric errors.}
  \label{mpfit}
\end{figure}

We determine the PNLF for the M87 halo as follows.  From the colour
and detection corrected PNLF, shown in the top panel of
Fig.~\ref{PNLF}, we subtract the expected contribution from Ly$\alpha$
emitters. Using the measured values of $\alpha_{2.5}$ for the M87 halo
and ICL PN population, we derive the fraction of PN in the M87 halo,
using
\begin{equation}
  \frac{\mathrm{N_{halo}}(m)}{\mathrm{N_{tot}}(m)} = 
   \frac{\alpha_{2.5,\mathrm{halo}}L_{\mathrm{halo}}}
      {\alpha_{2.5,\mathrm{halo}}L_{\mathrm{halo}}+\alpha_{2.5,\mathrm{ICL}}L_{\mathrm{ICL}}}
   = \frac{1}{1+3\frac{L_{\mathrm{ICL}}}{L_{\mathrm{halo}}}}
\end{equation}
where $L_{\mathrm{ICL}},\,L_{\mathrm{L_{halo}}}$ are given in
Table~\ref{tab_M31_M87} and
$\alpha_{2.5,\mathrm{ICL}}/\alpha_{2.5,\mathrm{halo}}=3$. We show the
PNLF of the halo of M87 in Fig.~\ref{mpfit}; error bars include
Poisson statistics and a 20\% cosmic variance of the Ly$\alpha$
density.

We now attempt to fit this PNLF with a generalised version of the
analytical formula reported in Eq.~\ref{PNLF_eq}. We continue to
assume that the bright cutoff near $M^*$ is invariant for different PN
populations, but we now allow for a free faint-end slope parameter
$c_2$, in addition to varying the parameter $c_1$ equivalent to sample
size. For the fit to the data we use robust non-linear least squares
curve fitting, i.e., the IDL routine mpfit \citep{markwardt09}, and
account for the photometric errors. We derive the following values for
the free parameters: $c_1=2017.2$, $c_2=1.17$ and $m^*=26.23$, for a
reduced $\chi^2 = 1.01$.  Fig.~\ref{mpfit} shows that this model is an
excellent fit to the empirical PNLF for the halo of M87.

The fitted value of $m^*=26.23$ corresponds to a nominal distance
modulus for M87 of $m-M = 30.74$ mag, or $D= 14.1$ Mpc. Since the
generalised PNLF formula has not been calibrated and our candidates
are not spectroscopically confirmed, we regard this as preliminary and
do not give an error on $m-M$. This value for the distance modulus is
$\sim 0.4$ mag brighter than that measured with the surface brightness
fluctuation method, $31.18\pm0.07$ mag \citep{mei07}, and with the tip
of the red giant branch, $31.12\pm0.14$ mag \citep{bird10}, which
correspond to distances of $17.2\pm0.5$ Mpc and $16.7\pm0.9$ Mpc.

\subsection{Comparison of the PNLF in the M87 halo and in 
the M31 bulge} 

Here we compare the observed properties of the PNLF in M87 with the
PNLF in the M31 bulge which was used to calibrate the analytical
formula by C98. The PN population of the M31 bulge was studied in
detail by \citet{ciardullo89}; it has similar depth to our PNLF,
i.e. it is complete 2.5 mag down the bright cutoff of the PNLF. Also,
the stellar population in the M31 has similar metallicity, colour and
stellar population age as the population in the inner $\sim 6\arcmin$
of M87; see Tab.~\ref{tab_M31_M87} for a complete list of the
parameters.

\begin{figure}
  \centering
  \includegraphics[width=9cm]{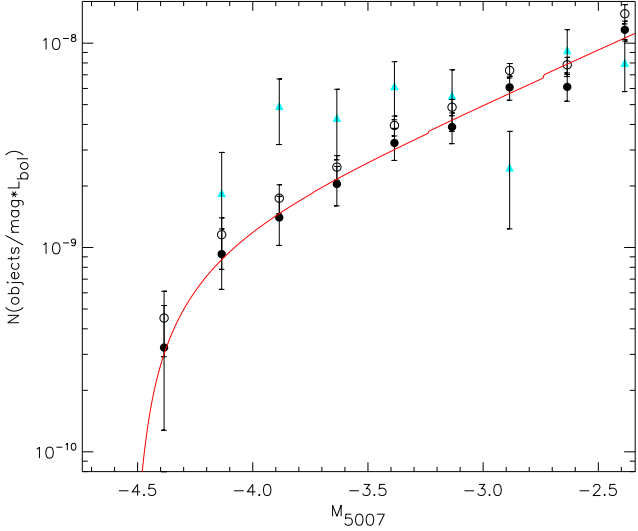}
  \caption{Luminosity-normalised PNLFs for the bulge of M31
    (triangles) and the halo of M87 (circles). Open circles and filled
    circles represent, respectively, the PNLF of the halo of M87
    before and after subtraction of the expected number of Ly$\alpha$
    contaminants in each bin. Data are binned into 0.25 mag
    intervals. M87 has a higher number of faint PNs per unit
    bolometric luminosity and its PNLF has a steeper slope towards
    faint magnitudes than M31.}\label{PNLF_M31_M87}
\end{figure}

We take the PN samples for M31 and the halo of M87 and normalise them
by the sampled luminosity, then correct for the distance modulus. The
results are shown in Fig.~\ref{PNLF_M31_M87}, where the PNLF for the
halo of M87 is shown before and after the subtraction of Ly$\alpha$
contaminants. For the points where the Ly$\alpha$ contribution is
subtracted the error bars take a $\sim 20\%$ cosmic variance of the
Ly$\alpha$ density (see Sect.~\ref{subsec:val}) into account. Within 1
mag of the bright cutoff, the M87 PN population has fewer PNs than the
PNLF of M31, i.e., the slope of the PNLF for the halo of M87 is
steeper towards fainter magnitudes.

In old stellar populations, one expects the PN central stars to be
dominated by low mass cores $M_{core} \le 0.55M_{\odot}$. Thus it is
plausible that the slope of the PNLF should be steeper than predicted
by modelling the fading of a uniformly expanding homogeneous gas
sphere ionised by a non-evolving star in a single PN
\citep{henize63}. The comparison between the PNLFs of M87 and M31 may
indicate that the M87 halo hosts a stellar population with a larger
fraction of low mass cores with respect to the M31 bulge.

From observations, there is further evidence that the faint end slope
of the luminosity function may depend on the parent stellar
population. \citet{ciardullo04} reported that star-forming systems,
like the disk of M33 and the Small Magellanic Cloud (SMC), have
shallower slopes $\sim1.5$ mags below the bright cutoff when compared
to older stellar populations such as NGC~5128 or M31. These shallower
slopes correspond to lower values of $c_2$ in Eq.~\ref{PNLF_eq}.

\begin{table*}
  \caption{Properties of the stellar populations in the bulge of M31 and in the halo of M87}
\label{tab_M31_M87}
 \centering
\begin{tabular}[h!]{c c c c c c c c}
     \hline\hline\\
     Name  & Distance  & L$_{\odot,\mathrm{bol}}$& (B-V)$_0$ & [Fe/H]$^a$ & $\alpha_{2.5}$&Age$^a$& Ref   \\
     \hline 
     & (Mpc)&  L$_{\odot,\mathrm{bol}}$& & &&Gyr&   \\
     \hline \\
     M31 (bulge)& 0.76& 6.5$\times 10^{9}$ &  0.95 & $\sim$ 0 & 16.3 $\times 10^{-9}$ &$ > 10$ &(1)\\
     M87 (halo)&14.5 &  6.2$\times 10^{10}$& 0.93&  $\sim$ 0&8.2 $\times 10^{-9}$&$>10$ &(2) \\
     \\
     \hline                  
   \end{tabular}
   \tablebib{
     (1) \citet{buzzoni06}; (2) This work.}
   \tablefoot{\tablefoottext{a} {Ratio [Fe/H] and Age from \citet{saglia10} (M31) 
   and \citet{liu05} (M87). M87 [Fe/H] and Age analysis covers the first 400\arcsec.}}
\end{table*}

\subsection{Comparison with previous PNLF distance 
measurements for M87 }
\label{PNLFoffset}

We already referred in Section~\ref{comparison_C98} to the work of C98
who carried out a PN survey in an area of $16\arcmin\times16\arcmin$
centred on M87, and investigated the properties of the PNLF. F03
studied the properties of ICPNs in the Virgo ICL, including an area of
$16\arcmin\times16\arcmin$ centred $14.8'$ north of M87. Both surveys
overlap with our current survey, but have a constant zero point offset
$\Delta=0.3$ mag relative to our measured magnitudes (see discussion
in Section~\ref{comparison_C98}).

\begin{figure}
  \centering
  \includegraphics[width=9cm]{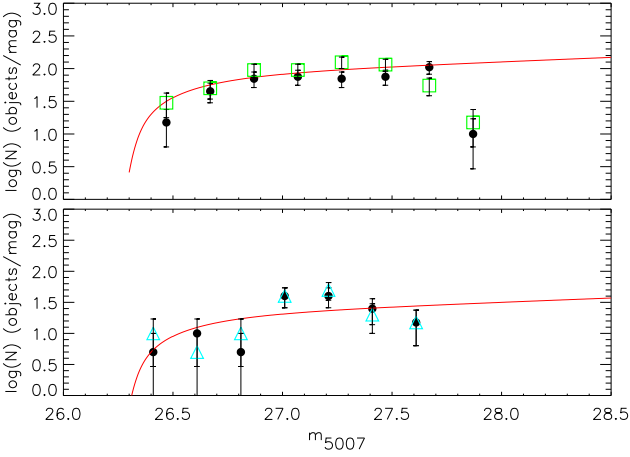}
  \caption{Luminosity functions of the matched PN samples from
      this survey (filled circles), from C98 (top panel, green
      squares), and from F03 (bottom panel, cyan triangles). The
      error bars show the 1 $\sigma$ uncertainty from counting
      statistics.  Data are binned into 0.2 mag intervals.  The solid
      red line represents the analytical PNLF model using a distance
      modulus of 30.8, convolved with photometric errors. The
      magnitude range is driven by the magnitude limit of the C98
      sample (top panel) and F03 sample (bottom panel). A zero point
      shift $\Delta=0.3$ mag relative to our sample was applied
      to both survey.}\label{PNLF_C98}
\end{figure}

Fig.~\ref{PNLF_C98} compares the PNLFs of the matched subsamples
common to our catalogue and those of C98 and F03, respectively.  A
residual zero point offset $\Delta=0.3$ mag has been applied to the
C98 and F03 subsamples. With this zero point shift, the empirical
PNLFs agree very well within one magnitude of the bright cutoff, and
are consistent with a distance modulus of 30.8. Without the zero point
shift, the distance modulus obtained from the PNLFs of C98 and F03
would be 30.5.  Thus the value obtained for our new M87 halo PN sample
(30.75), which is closer to other determinations
(Section~\ref{distmod}), indicates a systematic effect in the C98, F03
photometry, in the sense of brighter $m_{5007}$ in these samples.
This systematic effect can explain and resolve some of the tension
between the assumptions and results of C98, F03, and more recent
findings on the spatial distribution of ICPNs in the Virgo cluster and
the radial extension of the M87 halo, as discussed now.

C98 find that their empirical PNLF deviates from the analytical
formula and link the deviations with the presence of an ICPN
population uniformly distributed in the Virgo cluster volume. C98
assume that all PNs at distances larger than $R_{iso}> 4.8'$ from the
centre of M87 are ICPNs. Fitting a PNLF to this component, they derive
that the ICPN population must extend $4$ Mpc in front of M87; see
Fig.~8 and its figure caption in C98 for further details.  The depth
effect, which is equivalent to a brightening of the PNLF by foreground
ICPNs, places the near edge of the ICL population at a distance
modulus of 30.3, corresponding to a distance of 11 Mpc.

The assumptions of C98 on the membership of PNe at $R_{iso} > 4.8'$
from M87 to the ICL and on the spatial distribution of the ICL are not
supported by the results of the spectroscopic follow-up of
\citet{arnaboldi04} and the wide area ICPN survey by
\citet{castro09}. The spectroscopic follow-up by \citet{arnaboldi04}
showed that in the FCJ field of F03 at $14.8'$ from the M87 centre
87\% of the PNs are bound to the M87 halo, and only 13\% are
ICPNs. Thus the fraction of ICPNs in regions that are even closer to
the centre, as in the C98 $R_{iso} > 4.8'$, sample, will be down to a
few \%.  The wide area survey for ICPNs carried out by
\citet{castro09} shows that the ICL is associated with only the
densest regions of the Virgo cluster, $\sim0.4 $ Mpc around M87. Hence
the brightening of the PNLF expected due to foreground ICPNs is $<0.1$
mag. These results indicate that the C98 PNs at $R_{iso} > 4.8'$ are
bound to the M87 halo, with very limited contamination by ICL: thus
they are at the distance of M87 and the brightening of the C98 sample
is not caused by volume effects or a contamination by ICPNs. Similar
arguments apply to the F03 PN sample.

We therefore conclude that the magnitudes for the C98, F03 PNs sample
must be systematically too bright by $0.3$ mag, and that our new
photometry is reliable.

 \section{Summary and Conclusions}
\label{section6}

We carried out a deep survey for planetary nebulas (PNs) in two wide
fields of 34$\arcmin$ $\times$ 27$\arcmin$, covering the halo of the
cD galaxy NGC~4486 (M87) in the Virgo cluster with Suprime-Cam at the
Subaru telescope.  Both fields were imaged through a narrow-band
filter centred at the redshifted [OIII]$\lambda$5007 \AA\ emission and
a broad-band V filter.  The surveyed area covers the halo of M87 out
to a radial distance of 150 kpc. This is the largest survey so far for
PNs in the M87 halo in terms of number of detected PN candidates,
depth and area coverage.

PN candidates were identified via the on-off band technique on the
basis of automatic selection criteria using their narrow-band colours
and their two-dimensional light distributions.  The final photometric
catalogue contains 688 objects, with a magnitude range extending from
the apparent magnitude of the bright cutoff at Virgo distance down to
2.2 magnitude deeper.

We studied the radial number density profile of the PN candidates and
compared it with the V-band surface brightness profile of the stellar
light in the M87 halo. The logarithmic PN number density profile shows
good agreement with the $\mu_{\mathrm{V}}$ surface brightness profile
within 13$\arcmin$ from M87's centre, but flattens at large radii. At
the most distant point the difference to the prediction from the light
profile is 1.2 mags.  We investigated whether contributions from
background Ly$\alpha$ emitters at redshift $z=3.1$ can be responsible
for the flatter distribution.  By using the Ly$\alpha$ LF from
\citet{gronwall07} scaled to our effective surveyed volume, we
constrained the Ly$\alpha$ contribution to be $25\% \pm5\%$ of
the total sample over the whole surveyed area, and we concluded that
it cannot explain the factor 3 more sources in the outer regions that
are responsible for the flatter profile.

Stimulated by the early finding of \citet{doherty09} who determined
different luminosity specific PN numbers for the M87 halo light and
the ICL, we propose a two component model for the PN population, with
the ICL contributing a larger number of PNs per unit light. This
composite model is consistent with the observed flattening of the
logarithmic PN density profile when the luminosity specific PN numbers
are $\alpha_{2.5,\mathrm{halo}}=(1.10^{+0.17}_{-0.21})\times 10^{-8}$
N$_{\mathrm{PN}}$L$^{-1}_{\odot,\mathrm{bol}}$ and $
\alpha_{2.5,\mathrm{ICL}}=(3.29^{+0.60}_{-0.72})\times10^{-8}$
N$_{\mathrm{PN}}$L$^{-1}_{\odot,\mathrm{bol}}$, for the M87 halo and
ICL population respectively.

Because of the large magnitude range of this PN survey, we were able
to study the shape of the PNLF in detail and measured a steeper slope
at faint magnitudes than what is expected from the analytical formula
of \citet{ciardullo89}. The fit of the generalised PNLF formula
(Eq.~\ref{PNLF_eq}) to the empirical PNLF of the M87 halo population
gives faint-end slope of $c_2=1.17$ and a nominal distance modulus of
$30.74$.

The depth of the current survey allowed us to carry out a comparison
with the benchmark for PNLF studies, the M31 bulge.  The comparison of
the M87 halo PNLF with the M31 bulge PNLF, when both are normalised by
the sampled bolometric luminosity, shows that the former has fewer PNs
at bright magnitudes, and a steeper slope towards the faint end. The
steepening of the PNLF at fainter magnitudes is consistent with a
larger fraction of PNs with low mass cores. PN evolution models and
stellar population measurements at large radii will be needed to
understand the stellar population effects that shape the PNLF of the
M87 halo.

\begin{acknowledgements}
  We thank the on-site Subaru staff for their support and A. Aguerri
  and N. Castro-Rodriguez for helping with the routines that were used
  for the PNs catalogue extraction.  We are also thankful to G.Bono
  and A. Weiss for helpful discussions on stellar evolution in the AGB
  phase, and to the referee for her/his comments which improved the
  paper. AL thanks J. Elliott, B. Agarwal and G. Avvisati for the
  support and helpful comments on the paper. MAR would like to thank
  F. Makata for support in using the Suprime-Cam Data Reduction
  (sdfred) pipeline. LC acknowledges funding from the European
  Community's Seventh Framework Programme (FP7/2007-2013/) under grant
  agreement No 229517. This research has made use of the 2MASS
  catalogue data.
\end{acknowledgements}

\bibliographystyle{aa}
\bibliography{M87_halo}
\newpage
\newpage
\begin{table*}
  \caption{\label{table:density_profile}Number density and logarithm number density profile for the colour and spatial corrected sample of emission line objects.}
\centering          
\begin{tabular}{c c c c} 
\hline\hline\\
 R & $\frac{N_c}{A}$& $-2.5\log_{10}\left(\frac{N_c}{A}\right)+\mu_0$&C$(R)$\\
\hline\\
        (arcsec$^{1/4}$)& (N/arcsec$^2$) & mag/arcsec$^2$&\\
\hline \\
3.6&1.7 $\times 10^{-3}$  &22.9&  0.55\\
3.9&1.5 $\times 10^{-3}$  &23.1&  0.84\\
4.1&8.4 $\times 10^{-4}$  &23.7&  0.91\\
4.2&6.0 $\times 10^{-4}$  &24.0&  0.87\\
4.4&5.0 $\times 10^{-4}$  &24.2&  0.87\\
4.6&5.2 $\times 10^{-4}$  &24.2&  0.85\\
4.9&2.8 $\times 10^{-4}$  &24.9&  0.86\\
5.1&1.8 $\times 10^{-4}$  &25.3&  0.83\\
5.3&1.6 $\times 10^{-4}$  &25.5&  0.86\\
5.6&1.2 $\times 10^{-4}$  &25.8&  0.88 \\ 
5.8&1.2 $\times 10^{-4}$  &25.8&  0.88\\
6.0&7.4 $\times 10^{-5}$  &26.3&  0.88\\
6.4&5.1 $\times 10^{-5}$  &26.7&  0.91 \\
\\
\hline 
 \end{tabular}
 \tablefoot{The  term C$_{\mathrm{R}}$ is  the spatial completeness factor as
   function of the distance from M87's centre.}
\end{table*}

\begin{table*}
  \caption{\label{table:PNLF}PNLF for the colour and detection corrected sample of emission line objects.}
\centering          
\begin{tabular}{c c c c} 
\hline\hline\\
 m$_{5007}$ &$\log_{10}(N/bin)$& $Colour Completeness$ & $Detection Completeness$\\
\hline\\
 \hline \\
25.7&0.6&100\%&85\%\\
26.0&---&75\% &85\%\\
26.3&1.6&70\% &90\%\\
26.6&2.1&88\% &90\%\\
26.9&2.3&85\% &90\%\\
27.2&2.6&85\% &90\%\\
27.5&2.7&98\% &90\%\\
27.8&2.9&81\% &80\%\\
28.1&3.0&66\% &80\%\\
28.4&2.9&43\% &40\%\\

\\
\hline 
 \end{tabular}
\tablefoot{The  term $Colour Completeness$ represents the percentage of simulated PNs that we retrive with our colour selection, as function of the magnitude.\\
The term $Detection Completeness$ represents the recovery fraction of an input simulated PN population as function of magnitude.}
\end{table*}

\end{document}